\documentclass[aps,prd,10pt,nofootinbib,eqsecnum]{revtex4-2}
\usepackage[T1]{fontenc}
\usepackage[utf8]{inputenc}
\usepackage[a4paper]{geometry}
\usepackage{xcolor}
\usepackage{amsmath}
\usepackage{amssymb}
\usepackage{bm}
\usepackage{mathtools}
\usepackage{datetime2}
\usepackage{graphicx}
\usepackage{float}
\usepackage[unicode=true,pdfusetitle, bookmarks=true,bookmarksnumbered=false,bookmarksopen=false,
 breaklinks=false,pdfborder={0 0 1},backref=false,colorlinks=true]
 {hyperref}



\begin{document}
	
\title{Spatially covariant gravity with two degrees of freedom:\\
	A perturbative analysis up to cubic order}

\author{Yang Yu}%
	\affiliation{Center for Joint Quantum Studies and Department of Physics, School of Science, Tianjin University, Tianjin 300350, China}
    \affiliation{School of Physics and Astronomy, Sun Yat-sen University, Zhuhai 519082, China}
	
\author{Yu-Min Hu}%
    \affiliation{Institute for Gravitational Wave Astronomy, \\                              Henan Academy of Sciences, Zhengzhou, Henan 450046, China}

\author{Xian Gao}
\email[Corresponding author: ]{gaoxian@mail.sysu.edu.cn}

\affiliation{School of Physics, Sun Yat-sen University, Guangzhou 510275, China}
\affiliation{Guangdong Provincial Key Laboratory of Quantum Metrology and Sensing,
	Sun Yat-sen University, Zhuhai 519082, China}

\date{\today}

\begin{abstract}
	There has been considerable interest in constructing modified gravity theories that propagate only two degrees of freedom (DOFs), corresponding to the tensorial gravitational waves of general relativity. Within the framework of spatially covariant gravity (SCG), the conditions for obtaining 2-DOF theories can be derived from Hamiltonian constraint analysis, but it is generally difficult to translate those conditions into explicit SCG Lagrangians, especially when the Lagrangian depends nonlinearly on the extrinsic curvature. In this work, we adopt an alternative perturbative approach. We consider polynomial-type SCG Lagrangians up to $d=3$, where $d$ denotes the total number of derivatives in each monomial, and expand them around a cosmological background. By requiring the scalar mode to be eliminated up to cubic order in perturbations, we derive the corresponding conditions on the coefficient functions in the Lagrangian. We find five explicit Lagrangians that propagate only 2 DOFs up to cubic order in perturbations around a cosmological background. These theories therefore provide concrete candidate 2-DOF SCG models, at least at the perturbative level up to cubic order.
\end{abstract}

\maketitle

\section{Introduction}
For decades, modified gravity theories beyond general relativity (GR) have been studied extensively. Such investigations are motivated in part by cosmological phenomena that remain difficult to explain within GR, while they also provide a useful arena for examining the theoretical foundations of gravity from a broader field-theoretic perspective. In four-dimensional modified gravity, much of the traditional focus has been on introducing additional physical degrees of freedom (DOFs) to account for cosmological observations. More recently, however, increasing attention has been directed toward an alternative possibility: constructing modified gravity theories that preserve the same number of propagating DOFs as GR, namely, exactly two degrees of freedom (2-DOFs). This direction is further motivated by gravitational-wave observations from the LIGO-Virgo collaboration, which are highly consistent with pure tensor modes \cite{LIGOScientific:2017ycc,Takeda:2020tjj}, as well as by the absence of evidence for scalar dark matter \cite{LIGOScientific:2025ttj}. Moreover, 2-DOF theories naturally contain GR as a particular limit, making them compatible with current astronomical observations while still allowing for possible deviations in cosmological regimes.

As established by Lovelock's theorem \cite{Lovelock:1971yv,Lovelock:1972vz}, GR is the unique four-dimensional metric theory that respects spacetime diffeomorphism invariance and yields second-order field equations. This conclusion is also supported by the systematic construction of quadratic theories in the Fierz-Pauli framework, where a massless spin-2 field is endowed with the gauge symmetry induced by general coordinate transformations. To construct 2-DOF theories beyond GR, one must therefore evade Lovelock's theorem by relaxing some of its underlying assumptions. A particularly fruitful possibility is to break full spacetime diffeomorphism invariance, especially through the breaking of time-reparameterization symmetry. In this setting, gravity can be formulated in a $3+1$ decomposition, where the explicit spacetime split makes the time evolution and propagating modes transparent at the level of the equations of motion or in the Hamiltonian constraint structure. Theories that preserve only spatial diffeomorphism invariance are referred to as spatially covariant gravity (SCG) theories \cite{Gao:2014fra,Gao:2014soa,Gao:2018znj,Gao:2019lpz,Lin:2020nro}. Many previously known theories can be embedded into the SCG framework, including ghost condensation \cite{Arkani-Hamed:2003pdi}, the effective field theory of inflation \cite{Cheung:2007st,Gubitosi:2012hu}, and Ho\v{r}ava-Lifshitz gravity \cite{Horava:2009uw}. Through gauge-fixing and symmetry-restoration analyses, SCG theories are also closely related to scalar-tensor theories \cite{Gao:2020qxy,Gao:2020juc,Gao:2020yzr,Hu:2021bbo,Hu:2024hzo,Yu:2024sed}. In addition, gravitational-wave phenomenology and observational constraints in SCG have been studied extensively \cite{Fujita:2015ymn,Saitou:2016lvb,Gao:2019liu,Zhu:2022dfq,Zhu:2022uoq,Zhu:2023rrx,Yu:2024drx,Jiang:2025ysb,Feng:2026pcs}. Importantly, most known 2-DOF theories break temporal diffeomorphism invariance and thus fit naturally into the SCG framework.

The study of 2-DOF theories can be traced back to the cuscuton theory \cite{Afshordi:2006ad}, which propagates only two tensorial modes when the unitary gauge is allowed. The cuscuton theory has been extensively investigated \cite{Afshordi:2007yx,Bhattacharyya:2016mah,Boruah:2017tvg,Boruah:2018pvq,Quintin:2019orx,Bartolo:2021wpt,Maeda:2022ozc,Kohri:2022vst,HosseiniMansoori:2022xnq,Channuie:2023ddv,Mylova:2023ddj,Andrade:2023jvf} and further extended in \cite{Iyonaga:2018vnu,Iyonaga:2020bmm}. Other notable examples include the minimal theory of massive gravity \cite{DeFelice:2015hla,DeFelice:2015moy} and four-dimensional Einstein-Gauss-Bonnet gravity \cite{Glavan:2019inb}. Another important class is the minimally modified gravity (MMG) proposed in \cite{Lin:2017oow}, in which type-I MMG theories were first constructed by breaking spacetime diffeomorphism invariance and assuming an affine dependence on the lapse. Later, more general type-I MMG theories were obtained by rewriting GR through a general canonical transformation \cite{Aoki:2018zcv,Aoki:2018brq}. A similar strategy was then used to construct type-II MMG theories \cite{DeFelice:2020eju}.

Within the broader SCG framework, 2-DOF theories arising from SCG with a nonlinear dependence on the lapse \cite{Gao:2019twq} and on the velocity of the lapse function \cite{Lin:2020nro} have also been investigated. All of the above 2-DOF theories were constructed through Hamiltonian constraint analyses. In the Lagrangian formulation, one may in principle perform a Legendre transformation, derive the Hamiltonian, and then extract the constraint conditions from the corresponding Dirac algebra. Alternatively, one may work directly in the Hamiltonian formalism from the outset, which is often more economical but also less transparent. Examples of this latter strategy include MMG theories with an affine lapse dependence \cite{Mukohyama:2019unx} and 2-DOF theories obtained by introducing additional auxiliary constraints in phase space \cite{Yao:2020tur,Yao:2023qjd}. Related 2-DOF theories have also been widely studied in \cite{Carballo-Rubio:2018czn,Aoki:2020oqc,DeFelice:2020cpt,DeFelice:2020onz,DeFelice:2020prd,DeFelice:2020ecp,Sangtawee:2021mhz,Aoki:2021zuy,Iyonaga:2021yfv,DeFelice:2021xps,Pookkillath:2021gdp,Ganz:2022iiv,DeFelice:2022uxv,Jalali:2023wqh,Saito:2023bhn,Chakraborty:2023jek,Akarsu:2024qsi,Akarsu:2024eoo,Arora:2025msq,Ladeira:2026jne}, demonstrating their relevance for cosmology and compact stars.

Although a Hamiltonian construction can provide rigorous and general 2-DOF conditions, it is often difficult to translate those conditions into explicit solutions in the Lagrangian language, especially when the Lagrangian depends nonlinearly on the extrinsic curvature, i.e., on the velocity of the spatial metric. To address this difficulty, we adopt a fully Lagrangian approach. Since the number of DOFs can also be determined by applying constraint analysis directly to the equations of motion derived from a Lagrangian \cite{Mukunda:1974dr,Rothe:2010dzf}, we combine this idea with perturbation theory. A non-perturbative 2-DOF theory should propagate no scalar mode at any perturbative order around a cosmological background. Conversely, sufficient 2-DOF conditions can be obtained by requiring the scalar perturbation mode(s) to disappear order by order, which is achieved by appropriately tuning the coefficient functions in the Lagrangian. In practice, a finite set of such conditions may already be sufficient to eliminate all scalar mode(s) non-perturbatively. This strategy has previously been used to construct the extended cuscuton theory \cite{Iyonaga:2018vnu} and an SCG theory with the velocity of the lapse function propagating at most three degrees of freedom \cite{Gao:2019lpz}.

In \cite{Hu:2021yaq}, this perturbative method was applied to the construction of 2-DOF SCG. At the linear perturbation level, the conditions on the coefficients were derived for monomials up to $d=4$, where $d$ denotes the total number of derivatives in each monomial of the Lagrangian. However, eliminating the scalar mode at linear order, or equivalently at quadratic order in the perturbative Lagrangian, does not guarantee its absence at nonlinear orders. It is therefore necessary to extend the analysis beyond the linear level. In \cite{Hu:2021yaq}, only the $d=2$ Lagrangian was expanded to cubic order, and the additional conditions required to eliminate the scalar mode were obtained in that case. The present work continues that analysis by including the $d=3$ Lagrangian and determining the corresponding conditions on the coefficients such that the scalar mode is eliminated up to cubic order.

In this work, we apply the perturbative method around a Friedmann-Lema\^itre-Robertson-Walker (FLRW) background to a polynomial-type SCG theory up to $d=3$, which propagates at most three degrees of freedom. By expanding the action up to cubic order in perturbations, we derive the necessary conditions for eliminating the propagating scalar mode order by order. By tuning the functional forms of the coefficients so that these 2-DOF conditions are satisfied, we identify several candidate Lagrangians. These theories propagate only 2 DOFs, at least up to cubic order in perturbations.

This paper is organized as follows. In Sec. \ref{sec:2}, we briefly review the basic formalism of spatially covariant gravity and present the explicit polynomial-type SCG Lagrangian with 3 degrees of freedom up to $d=3$, which serves as the starting point of our construction. We also summarize the sufficient conditions for eliminating the extra scalar mode in the quadratic perturbation action. In Sec. \ref{sec:3}, we expand the action to cubic order around a homogeneous and isotropic cosmological background and derive the 2-DOF conditions by requiring the scalar mode to be completely removed at each perturbative order. This procedure allows us to construct 2-DOF theories, valid at least up to cubic order, by appropriately tuning the coefficients from the outset, and it eventually leads to several specific Lagrangians. In Sec. \ref{sec:4}, we compare our results with other theories. Finally, Sec. \ref{sec:5} contains our conclusions. Throughout this paper, we use the unit $8\pi G=1$ and the metric signature $\{-,+,+,+\}$. Latin indices $\{i,j,k...\}$ denote spatial coordinates.

\section{From SCG to 2-DOF theories: perturbation analysis} \label{sec:2}
This section briefly reviews spatially covariant gravity (SCG) theories and sets up the perturbative framework used in the rest of the paper. In particular, we focus on a polynomial-type SCG action that propagates three degrees of freedom (DOFs), which will serve as the starting point for our subsequent construction. We then outline the perturbative method used to derive the degeneracy conditions for 2-DOF theories and summarize the relevant results obtained in previous work.

\subsection{Spatial covariant gravity and polynomial-type model with 3 DOFs}
\label{subsec:2.1}

Spatial covariant gravity (SCG) is, in general, a theoretical framework that respects three-dimensional spatial diffeomorphisms. The Arnowitt-Deser-Misner (ADM) variables are naturally suited to its foliated spacetime interpretation. The action of SCG is built from spatial geometrical quantities and their derivatives. The intrinsic geometry on each hypersurface is described by the spatial metric, while the normal vector governs the evolution of the slices. The general action for SCG takes the form
\begin{align}
	S=\int \mathrm{d}t\mathrm{d}^{3}x\:N\sqrt{h}\mathcal{L}\left(t,N,h_{ij},R_{ij},\nabla_{i},\pounds_{\boldsymbol{n}}\right).
\end{align}
Here $N$ is the lapse function, $h_{ij}$ is the spatial metric, $R_{ij}$ is the spatial Ricci tensor, $\nabla_{i}$ is the covariant derivative compatible with the spatial metric, and $\pounds_{\boldsymbol{n}}$ is the Lie derivative with respect to the normal vector $\boldsymbol{n}$. Since SCG theories preserve spatial diffeomorphisms but not full spacetime diffeomorphisms, the lapse function $N$ can enter the Lagrangian nonlinearly, reflecting the breaking of time-reparameterization invariance. By contrast, the shift vector $N_{i}$ merely encodes the gauge freedom associated with the choice of spatial coordinates.

If no further restrictions are imposed on this action, the presence of higher-order Lie derivatives strongly suggests the appearance of ghostlike DOFs, since the Lie derivative plays the role of temporal evolution in SCG theory. To avoid such pathological behavior and construct a healthy theory, an effective and convenient assumption is that first-order Lie derivatives act only on the spatial metric. This serves as a sufficient, though not necessary, condition for obtaining a 3-DOF SCG theory (see \cite{Gao:2018znj} for the 3-DOF condition with a dynamical lapse function). However, once temporal diffeomorphism invariance is broken, the lapse function $N$ enters freely as a fundamental variable, much like the spatial metric $h_{ij}$, thereby introducing an additional scalar mode into the theory.

Consequently, the action for a class of parity-preserving, 3-DOF SCG theories generally takes the form
\begin{align}
	S & =\int\text{d}t\text{d}^{3}x\:N\sqrt{h}\mathcal{L}(t,N,h_{ij},K_{ij},R_{ij},\nabla_{i}),\label{eq:general SCG action}
\end{align}
where $K_{ij}$ is the extrinsic curvature defined by 
\begin{align}
	K_{ij} &\coloneqq \frac{1}{2}\pounds_{\boldsymbol{n}}h_{ij}=\frac{1}{2N}(\partial_{t}h_{ij}-\nabla_{i}N_{j}-\nabla_{j}N_{i}).
\end{align}
It represents the first order of the Lie derivative of the spatial metric. In Ref. \cite{Gao:2014fra}, a detailed Hamiltonian analysis of the SCG action \eqref{eq:general SCG action} was carried out, and it was shown that, without any additional condition, the theory propagates at most 3 DOFs: one scalar mode and two tensor modes.

In this work, we focus on a specific Lagrangian built from polynomial monomials within the SCG framework. Each such monomial can be recast into the form of a generally covariant scalar-tensor (GST) theory through a gauge-recovering procedure, namely, the ``Stuckelberg trick'' \cite{Gao:2020qxy,Gao:2020yzr,Hu:2021bbo}. Within the resulting GST formalism, each monomial is characterized by a set of integers $(c_{0};d_{2},d_{3})$, and the total number of derivatives is
\begin{align}
	d & =\sum_{n=0}[(n+2)c_{n}+(n+1)d_{n+2}],
\end{align}
where $c_{n}$ and $d_{n}$ denote the numbers of the $n$-th order covariant derivatives of the spacetime Riemann curvature tensor and the scalar field, respectively. Since each SCG monomial corresponds, through gauge recovering, to linear combinations of monomials in GST theories with the same $d$ (see \cite{Gao:2020qxy,Gao:2020juc,Gao:2020yzr} for details), the parameter $d$ provides an overall derivative ``order'' that facilitates a transparent classification between the two frameworks. For example, after gauge recovering and the elimination of dangerous terms, the Horndeski Lagrangians $\mathcal{L}^{\text{H}}_4$ and $\mathcal{L}^{\text{H}}_5$ in the unitary gauge arise as specific degenerate combinations of SCG polynomials at $d=2$ and $d=3$, respectively (see Ref. \cite{Hu:2021bbo} for further details). Motivated by this correspondence with GST theories, we restrict our attention to explicit SCG models up to $d=3$.

The general polynomial-type SCG action up to $d=3$ is \cite{Hu:2021yaq}
\begin{align}
	S & =\int\text{d}t\text{d}^{3}xN\sqrt{h}(\mathcal{L}^{(0)}+\mathcal{L}^{(1)}+\mathcal{L}^{(2)}+\mathcal{L}^{(3)}),\label{eq:monomial-type SCG action up to d=3}
\end{align}
with
\begin{align}
	\mathcal{L}^{(0)} & =c_{1}^{(0;0,0)},\label{eq:L0}
\end{align}
\begin{align}
	\mathcal{L}^{(1)} & =c_{1}^{(0;1,0)}K,\label{eq:L1}
\end{align}
\begin{align}
	\mathcal{L}^{(2)} & =c_{1}^{(0;2,0)}K^{ij}K_{ij}+c_{2}^{(0;2,0)}a^{i}a_{i}+c_{3}^{(0;2,0)}K^{2}+c_{1}^{(1;0,0)}R,\label{eq:L2}
\end{align}
and 
\begin{align}
	\mathcal{L}^{(3)} & =c_{1}^{(0;3,0)}K_{ij}K^{jk}K_{k}^{i}+c_{2}^{(0;3,0)}K_{ij}a^{i}a^{j}+c_{3}^{(0;3,0)}K_{ij}K^{ij}K+c_{4}^{(0;3,0)}Ka^{i}a_{i}+c_{5}^{(0;3,0)}K^{3}\nonumber \\
	& \ \ \ \ +c_{1}^{(0;1,1)}K_{ij}\nabla^{i}a^{j}+c_{2}^{(0;1,1)}K\nabla_{i}a^{i}\nonumber \\
	& \ \ \ \ +c_{1}^{(1;1,0)}R^{ij}K_{ij}+c_{2}^{(1;1,0)}RK.\label{eq:L3}
\end{align}
Here $a_{i}\equiv\nabla_{i}\ln N$ is the acceleration, and the coefficients $c_{m}^{(c_{0};d_{2},d_{3})}$ are general functions of time $t$ and the lapse function $N$.

\subsection{Perturbation method and degeneracy conditions}
\label{subsec:2.2}

In \cite{Gao:2019twq}, the Hamiltonian constraint analysis in phase space was employed to derive 2-DOF conditions, with the $d=2$ SCG theory presented as an explicit example. Although constraint analysis can provide strict and universal 2-DOF conditions for general theories, its practical implementation is often highly involved, especially when the relation between momentum and velocity is nonlinear.

In this work, we instead adopt cosmological perturbation theory as a practical method for deriving explicit 2-DOF conditions. We take the polynomial-type SCG theory with 3 DOFs as the starting action and aim to construct 2-DOF theories by eliminating the single scalar DOF. In general, this scalar DOF appears as a propagating mode around a given background. Our strategy is therefore to eliminate the scalar mode order by order in scalar perturbations by imposing suitable conditions on the coefficients. This method has already been used successfully in the extended cuscuton theory \cite{Iyonaga:2018vnu} and in classes of SCG theories \cite{Gao:2019lpz,Hu:2021yaq,Wang:2024hfd}. Once the degeneracy conditions for the scalar mode at a given order have been extracted, we return to the original SCG model and fine-tune the coefficients so as to satisfy these conditions from the outset. In this way, the scalar mode is removed at each order of the perturbative action, and the resulting SCG model propagates only 2 DOFs. Since the theory under consideration is already known to propagate at most three DOFs, it is sufficient to impose degeneracy conditions that eliminate the single scalar DOF.

One may wonder whether such a perturbative program requires an analysis at infinitely many orders and is therefore impractical. In fact, the number of coefficients appearing in the SCG Lagrangian is finite, as is the number of corresponding monomials. Consequently, only a finite number of conditions is expected to be required to eliminate the scalar mode completely, even in a nonperturbative sense. These conditions can be regarded as a set of differential equations for the coefficients, and they may already be fixed by requiring the perturbative Lagrangian to be degenerate up to some finite order.

SCG theory encompasses a broad class of modified gravity theories that respect only spatial diffeomorphism invariance. It is therefore naturally formulated in ADM variables within a $3+1$ decomposition. We consider a Friedmann-Lema\^itre-Robertson-Walker (FLRW) cosmological background and describe perturbations as deviations of the ADM variables from their background values. For the purposes of this work, we restrict attention to scalar perturbations,
\begin{align}
	N =\bar{N}e^{A}, \qquad N_{i}  =\bar{N}a\partial_{i}B, \qquad h_{ij} & =a^{2}e^{2\zeta}\delta_{ij},
	\label{eq:perturbed}
\end{align}
with the scale factor $a(t)$, the background lapse $\bar{N}(t)$, and the scalar perturbation variables $A$, $B$, and $\zeta$. In general, these theories do not respect time-reparameterization symmetry, and there is therefore no reason to set $\bar{N}$ to a particular value by hand. According to the perturbative order, the action can be expanded as
\begin{align}\label{eq:perturbation in orders}
	S_{\text{SCG}}  =\int\text{d}t\text{d}^{3}x\bar{N}a^{3}(\mathcal{L}_{0}+\mathcal{L}_{1}+\mathcal{L}_{2}+\mathcal{L}_{3}+\cdots),
\end{align}
where the subscript $n$ in $\mathcal{L}_n$ denotes the order of the perturbed Lagrangian.

Since the lapse function $N$ is nondynamical and the shift vector $N^i$ cannot be treated as a free variable if spatial gauge symmetry is to be maintained, their corresponding scalar perturbations, namely $A$ and $B$, play the role of auxiliary variables. Their constraint equations can therefore be solved in terms of $\zeta$. Substituting these solutions back into the perturbed action yields an effective action depending only on the single scalar variable $\zeta$\footnote{One may refer to \cite{Wang:2013zva} for a nice introduction to the perturbation methods.}. The 2-DOF conditions up to the $n$-th order of the perturbative action are then equivalent to requiring that all coefficients of the terms contributing to the dynamics of $\zeta$ up to the $n$-th order vanish.

\subsection{2-DOF theories up to quadratic perturbation action}

The starting point is a subclass of the monomial-type SCG action \eqref{eq:monomial-type SCG action up to d=3}, namely,
\begin{align}
	S & =\int\text{d}t\text{d}^{3}xN\sqrt{h}(\mathcal{L}^{(0)}+\mathcal{L}^{(2)}+\mathcal{L}^{(3)}-\Lambda),\label{eq:action of d=0, d=2 and d=3}
\end{align}
where $\Lambda$ is the cosmological constant introduced as a function of time.
Compared with the model in \cite{Hu:2021yaq}, we here include $\mathcal{L}^{(0)}$ as an effective $k$-essence term.
In this subsection, we recapitulate the results of \cite{Hu:2021yaq}, in which the scalar perturbation theory of (\ref{eq:action of d=0, d=2 and d=3}) (in the absence of $\mathcal{L}^{(0)}$) was studied in detail.

For later convenience, we decompose the extrinsic curvature into its trace and traceless parts, $K_{ij}=\hat{K}_{ij}+\frac{1}{3}h_{ij}K$ with $\hat{K}_{ij}h^{ij}\equiv 0$. The Lagrangians then take the form
\begin{align}
	\mathcal{L}^{(0)}=a_1
\end{align}
with
\begin{align}
	a_1=c_{1}^{(0;0,0)},
\end{align}
\begin{align}  \mathcal{L}^{(2)}&=\omega_{2}\hat{K}^{ij}\hat{K}_{ij}+\tilde{b}_{2}a^{i}a_{i}+\frac{1}{3}b_{2}K^{2}+h_{2}R
\end{align}
with
\begin{align}
	&\omega_{2}=c_{1}^{(0;2,0)},\\
	&b_{2}=c_{1}^{(0;2,0)}+3c_{3}^{(0;2,0)},\\
	&\tilde{b}_{2}=c_{2}^{(0;2,0)},\\
	&h_{2}=c_{1}^{(1;0,0)},
\end{align}
and
\begin{align}  \mathcal{L}^{(3)}&=c_{1}^{(0;3,0)}\hat{K}_{ij}\hat{K}^{jk}\hat{K}_{k}^{i}+c_{2}^{(0;3,0)}\hat{K}_{ij}a^{i}a^{j}+\omega_{3}\hat{K}_{ij}\hat{K}^{ij}K+\frac{1}{3}\tilde{b}_{3}Ka^{i}a_{i}+\frac{1}{9}b_{3}K^{3}\nonumber \\
	&\ \ \ \ +c_{1}^{(0;1,1)}\hat{K}_{ij}\nabla^{i}a^{j}+\frac{1}{3}f_{3}K\nabla_{i}a^{i}\nonumber \\
	&\ \ \ \ +c_{1}^{(1;1,0)}R^{ij}\hat{K}_{ij}+\frac{1}{3}J_{3}RK
\end{align}
with
\begin{align} 
	&\omega_{3}=c_{1}^{(0;3,0)}+c_{3}^{(0;3,0)},\\
	&\tilde{b}_{3}=c_{2}^{(0;3,0)}+3c_{4}^{(0;3,0)},\label{b3tld}\\
	&b_{3}=c_{1}^{(0;3,0)}+3c_{3}^{(0;3,0)}+9c_{5}^{(0;3,0)},\\
	&f_{3}=c_{1}^{(0;1,1)}+3c_{2}^{(0;1,1)},\\
	&J_{3}=c_{1}^{(1;1,0)}+3c_{2}^{(1;1,0)}.
\end{align}
Here we have redefined a set of coefficients following the conventions of \cite{Hu:2021yaq}. We stress that these Lagrangians are equivalent to those in (\ref{eq:action of d=0, d=2 and d=3}), and that no conditions have yet been imposed on the coefficients. To ensure that GR is contained in our theory and that the tensor modes propagate, we restrict our analysis to the case with $\omega_{2} \neq 0$ and $b_{2} \neq 0$.

After substituting the perturbed quantities \eqref{eq:perturbed} and expanding the scalar perturbations up to quadratic order, we collect the action according to the perturbative order as in \eqref{eq:perturbation in orders}. The variation of the first-order perturbation action yields two independent background evolution equations:
\begin{align}
	0= & -\Lambda+a_{1}-b_{2}(3H^{2}+2\dot{H})-6Hb_{3}(H^{2}+\dot{H})-2H\dot{b}_{2}-3H^{2}\dot{b}_{3},\label{eq:bgeq1}\\
	0= & -\Lambda+(a_{1}^{\prime}+a_{1})+3H^{2}(b_{2}^{\prime}-b_{2})+3H^{3}(b_{3}^{\prime}-2b_{3}),\label{eq:bgeq2}
\end{align} 
with the Hubble parameter $H\coloneqq\frac{\dot{a}}{a}$. Throughout this paper, a dot and a prime denote derivatives with respect to time $t$ and the lapse function $N$, respectively, and are defined as
\begin{align}
	\dot{X}  & =\frac{1}{\bar{N}}\frac{\partial X}{\partial t}, \qquad \ddot{X}  =\frac{1}{\bar{N}}\frac{\partial}{\partial t}\Big(\frac{1}{\bar{N}}\frac{\partial X}{\partial t}\Big),\\
	f^{\prime} & =\bar{N}\Big(\frac{\partial f}{\partial N}\Big)_{N=\bar{N}}, \qquad f^{\prime\prime}  =\bar{N}^{2}\Big(\frac{\partial^{2}f}{\partial N^{2}}\Big)_{N=\bar{N}},
\end{align}
and so on. The quadratic Lagrangian for scalar perturbations takes the schematic form
\begin{align}
	\mathcal{L}_{2}(\zeta,A,B) & \simeq\dot{\zeta}\hat{\mathcal{C}}_{\dot{\zeta}\dot{\zeta}}\dot{\zeta}+\dot{\zeta}\hat{\mathcal{C}}_{\dot{\zeta}A}A+\dot{\zeta}\hat{\mathcal{C}}_{\dot{\zeta}B}B+A\hat{\mathcal{C}}_{AA}A+A\hat{\mathcal{C}}_{AB}B+B\hat{\mathcal{C}}_{BB}B\nonumber \\
	& \ \ \ \ +\zeta\hat{\mathcal{C}}_{\zeta\zeta}\zeta+\zeta\hat{\mathcal{C}}_{\zeta A}A+\zeta\hat{\mathcal{C}}_{\zeta A}B,
\end{align}
where the coefficients $\hat{\mathcal{C}}_{\dot{\zeta}\dot{\zeta}}$, $\hat{\mathcal{C}}_{\dot{\zeta}A}$, etc. generally depend on $t$ and may contain spatial derivatives.
In general, varying the action with respect to $A$ and $B$ yields two independent constraint equations. These can be solved formally and expressed in terms of $\zeta$. Substituting the two solutions back into the quadratic Lagrangian then leads to an effective Lagrangian $\mathcal{L}_{2}(\zeta)$ that depends only on $\zeta$.

The next step is to determine the degeneracy condition for the scalar perturbation $\zeta$, which is equivalent to requiring that the kinetic term of $\zeta$ vanish. After some manipulations, one finds that the condition for $\zeta$ to be nondynamical is $\Delta=0$, where \cite{Hu:2021yaq}
\begin{equation}
	\Delta\coloneqq\hat{\mathcal{C}}_{\dot{\zeta}\dot{\zeta}}(4\hat{\mathcal{C}}_{BB}\hat{\mathcal{C}}_{AA}-\hat{\mathcal{C}}_{AB}^{2})+\hat{\mathcal{C}}_{AB}\hat{\mathcal{C}}_{\dot{\zeta}A}\hat{\mathcal{C}}_{\dot{\zeta}B}-\hat{\mathcal{C}}_{AA}\hat{\mathcal{C}}_{\dot{\zeta}B}^{2}-\hat{\mathcal{C}}_{BB}\hat{\mathcal{C}}_{\dot{\zeta}A}^{2} .\label{eq:Delta}
\end{equation}
At first sight, (\ref{eq:Delta}) is a single constraint equation. However, since it contains spatial derivatives (more precisely, $\partial^{2}\coloneqq\partial_i\partial^i$) and depends on the Hubble parameter $H$, one must require the coefficients of terms with different powers of $\partial^{2}$ and $H$ to vanish separately. Under this requirement, a set of solutions is found to be \cite{Hu:2021yaq}
\begin{align}
	a_{1}&=\frac{B_{1}}{N}+B_{2},\\
	b_{2} & =\frac{C_{1}N}{1+C_{2}N},\label{eq:solution for b2}\\
	b_{3} & =\frac{D_{1}N^{2}}{(1+C_{2}N)^{2}},\label{eq:solution for b3}\\
	\tilde{b}_{3} &=\frac{C_{2}Nf_{3}}{1+C_{2}N}+f_{3}^{\prime}.\label{eq:solution for bt3}
\end{align}
Here $B_{1},B_{2},C_{1},C_{2}$, and $D_{1}$ are general functions of time\footnote{This is the same to all the integration constants (i.e., $C_{i}$'s and $D_{i}$'s) in the following.}. The distinct parts of the solutions are further classified into two cases:
\begin{itemize}
	\item Case I: $f_{3}=\tilde{f}_{3}=0$ and thus 
	\begin{align}
		&\tilde{b}_{3}=0, \label{caseI_b3t}
	\end{align}
	\item Case II: 
\end{itemize}
\begin{align}
	\omega_{2} & =-\frac{b_{2}(f_{3}-3\tilde{f}_{3})^{2}}{2f_{3}^{2}}, \label{case2omg2}\\
	\omega_{3} & =-\frac{b_{3}(f_{3}-3\tilde{f}_{3})^{2}}{2f_{3}^{2}},\label{eq:solution for omega3}
\end{align}
with $f_{3}\neq0$ and $f_{3}-3\tilde{f}_{3}\neq 0 $ where $\tilde{f}_{3}\coloneqq c_{1}^{(0;1,1)}+3c_{2}^{(0;1,1)}$.

With the above solutions for the coefficients, the quadratic Lagrangian for the perturbations propagates no scalar DOF.
In \cite{Hu:2021yaq}, the constraints on the coefficients required to eliminate the scalar DOF in the cubic Lagrangian were studied, but only for $d=2$, namely, in the absence of $\mathcal{L}^{(3)}$.
The purpose of the present work is to further suppress the scalar mode in the cubic Lagrangian and, in particular, to extend the analysis of \cite{Hu:2021yaq} by constraining the coefficients of the $d=3$ terms.
To this end, a higher-order perturbation analysis is necessary.
In the next section, we will expand the perturbed action to cubic order and require the complete disappearance of any dynamical scalar mode, thereby deriving further conditions that restrict the functional form of the Lagrangian.

\section{Degeneracy conditions in the cubic order of perturbations} \label{sec:3}

In a theory with no propagating scalar mode, the kinetic terms of the scalar perturbations must vanish, or become degenerate, at every order in perturbation theory. Although one can derive conditions at quadratic order and tune the coefficients accordingly, those conditions are not, in general, sufficient to guarantee the absence of the scalar mode at higher orders. 
For the monomial-type SCG theory with $d=2$, namely the action \eqref{eq:monomial-type SCG action up to d=3} with $\mathcal{L}^{(3)}=0$, the quadratic analysis constrains only the coefficients $b_2$ and $\tilde{b}_2$. The additional conditions on $\omega_2$ and $h_2$ arise only from the cubic perturbation Lagrangian (see Sec. V of \cite{Hu:2021yaq} for details). It is therefore necessary to extend the analysis to cubic order in perturbations for the present model.

We now turn to the main task of this work. The 2-DOF conditions for the action \eqref{eq:action of d=0, d=2 and d=3} up to quadratic order in perturbations were derived in \cite{Hu:2021yaq}. There it was shown that the scalar mode at this order can be avoided in two distinct classes of coefficient functions, given in (\ref{caseI_b3t})-(\ref{eq:solution for omega3}). In each subsection below, we begin from the corresponding action obtained in that previous analysis. For completeness, we also present the associated linear-order solutions for $A$ and $B$. 

Following the same strategy as in the quadratic analysis, we substitute the solutions for $A$ and $B$ into the cubic Lagrangian $\mathcal{L}_{3}(\zeta,A,B)$ and thereby obtain an effective cubic Lagrangian $\mathcal{L}_{3}(\zeta)$ for the single scalar variable $\zeta$. Starting from the quadratic-order conditions, we then impose further restrictions on the coefficient functions so as to eliminate the propagating scalar mode completely. This procedure leads to several candidate 2-DOF Lagrangians, which constitute the main results of this paper.

\subsection{Case I}
\label{subsec:3.1}

For case I, the action takes the form
\begin{align}
	S^{\text{I}} & =\int\text{d}t\text{d}^{3}x\:N\sqrt{h}\Big[\omega_{2}K^{ij}K_{ij}+\frac{1}{3}\Big(\frac{C_{1}N}{1+C_{2}N}-\omega_{2}\Big)K^{2}+h_{2}{}^{3}\!R+\Big(\frac{B_{1}}{N}+B_{2}\Big)\nonumber \\
	& \ \ \ \ +c_{1}^{(0;3,0)}K_{ij}K^{jk}K_{k}^{i}-3c_{4}^{(0;3,0)}K_{ij}a^{i}a^{j}+(\omega_{3}-c_{1}^{(0;3,0)})K_{ij}K^{ij}K+c_{4}^{(0;3,0)}Ka^{i}a_{i}\nonumber \\
	& \ \ \ \ +\frac{1}{9}\Big(\frac{D_{1}N^{2}}{(1+C_{2}N)^{2}}-c_{1}^{(0;3,0)}-3c_{3}^{(0;3,0)}\Big)K^{3}\nonumber \\
	& \ \ \ \ +c_{1}^{(1;1,0)}R^{ij}K_{ij}+\frac{1}{3}(J_{3}-c_{1}^{(1;1,0)})RK\Big].
\end{align}
By expanding $S^{\mathrm{I}}$ up to quadratic and cubic order in the perturbations $A$, $B$, and $\zeta$, we obtain the corresponding Lagrangians $\mathcal{L}^{\mathrm{I}}_{2}(A,B,\zeta)$ and $\mathcal{L}^{\mathrm{I}}_{3}(A,B,\zeta)$.
From $\mathcal{L}^{\mathrm{I}}_{2}$, the linear-order solutions for the auxiliary variables $A$ and $B$ in terms of $\zeta$ are
\begin{align}
	A&=\frac{1+C_{2}\bar{N}}{C_{2}H\bar{N}}\dot{\zeta}+\Big(\frac{D_{1}(c_{1}^{(1;1,0)}+2J_{3})(1+C_{2}\bar{N})\zeta}{2C_{2}(C_{1}+C_{1}C_{2}\bar{N}+3D_{1}H\bar{N})a^{2}(\omega_{2}+3H\omega_{3})}\nonumber\\&\ \ \ \ +\frac{(h_{2}+h_{2}^{\prime}+HJ_{3}^{\prime})(1+C_{2}\bar{N})\zeta\bigl(C_{1}\bar{N}(1+C_{2}\bar{N})^{2}+2(1+C_{2}N)^{3}(\omega_{2}+3H\omega_{3})\bigr)}{3C_{2}{}^{2}H^{2}\bar{N}^{3}(C_{1}+C_{1}C_{2}\bar{N}+3D_{1}H\bar{N})a^{2}(\omega_{2}+3H\omega_{3})}\nonumber\\&\ \ \ \ +\frac{\bigl(C_{1}C_{2}(c_{1}^{(1;1,0)}+2J_{3})+6D_{1}(h_{2}+h_{2}^{\prime}+HJ_{3}^{\prime})\bigr)(1+C_{2}\bar{N})^{2}\zeta}{6C_{2}{}^{2}H\bar{N}(C_{1}+C_{1}C_{2}\bar{N}+3D_{1}H\bar{N})a^{2}(\omega_{2}+3H\omega_{3})}\Big)\partial^{2}\zeta,\\B&=-\Big(\frac{c_{1}^{(1;1,0)}+2J_{3}}{2a(\omega_{2}+3H\omega_{3})}+\frac{(h_{2}+h_{2}^{\prime}+HJ_{3}^{\prime})(1+C_{2}\bar{N})}{C_{2}H\bar{N}a(\omega_{2}+3H\omega_{3})}\Big)\zeta,
\end{align}
where we have assumed the non-degeneracy condition
\begin{equation}
	C_2\left( \omega_2+3H\omega_3\right)(C_{1}+C_{1}C_{2}\bar{N}+3D_{1}H\bar{N})\neq 0,
\end{equation}
so that the auxiliary variables can indeed be solved. Substituting these expressions for $A$ and $B$ into $\mathcal{L}^{\mathrm{I}}_{3}(A,B,\zeta)$ yields the effective cubic Lagrangian for the single scalar variable $\zeta$. In what follows, we denote $\mathcal{L}^{\mathrm{I}}_{3}\equiv \mathcal{L}^{\mathrm{I}}_{3}(\zeta)$ for brevity.

Since the original SCG theory does not contain higher time derivatives, $\zeta$ appears with at most one time derivative. Accordingly, at most three time derivatives can appear in the cubic perturbation Lagrangian. If one were to use integrations by parts in time to reduce terms with three time derivatives, the resulting degeneracy conditions would explicitly involve the time evolution. However, a general 2-DOF theory should not depend on the particular realization of the scalar field, namely on its identification with time in the unitary gauge. The solutions are expressed in terms of the lapse function, while the scalar field (time) is encoded as an integration constant and appears only implicitly as the argument of an implicit function. For this reason, we perform integrations by parts only with respect to spatial derivatives. 

\begin{figure}[H]
\includegraphics[scale=0.27]{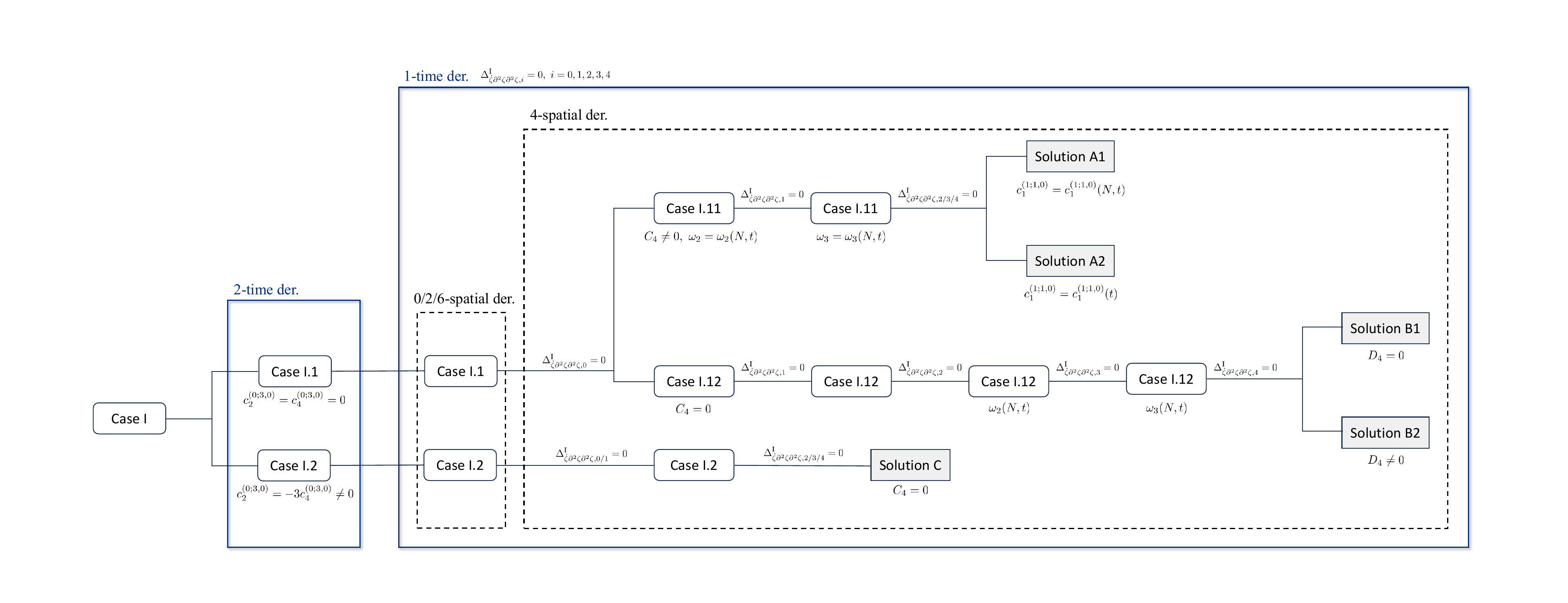}
\caption{Flowchart for Case I.}
\end{figure}

It is then convenient to classify the resulting conditions according to the number of time derivatives.

\begin{itemize}
	\item{\textbf{Terms with three time derivatives}}
\end{itemize}

After integrations by parts over spatial derivatives, all terms with three time derivatives cancel identically.

\begin{itemize}
	\item{\textbf{Terms with two time derivatives}}
\end{itemize}
The terms containing two time derivatives in the cubic Lagrangian take the form
\begin{align}
	\mathcal{L}_{3,t2}^{\text{I}} & \simeq\hat{\mathcal{C}}_{\dot{\zeta}^{2}\partial^{2}\zeta}^{\text{I}}\dot{\zeta}^{2}\partial^{2}\zeta+\hat{\mathcal{C}}_{\partial_{i}\dot{\zeta}\partial^{i}\dot{\zeta}\partial^{2}\zeta}^{\text{I}}\partial_{i}\dot{\zeta}\partial^{i}\dot{\zeta}\partial^{2}\zeta+\hat{\mathcal{C}}_{\partial_{i}\dot{\zeta}\partial_{j}\dot{\zeta}\partial^{i}\partial^{j}\zeta}^{\text{I}}\partial_{i}\dot{\zeta}\partial^{i}\dot{\zeta}\partial^{2}\zeta
\end{align}
with
\begin{align}
	\hat{\mathcal{C}}_{\dot{\zeta}^{2}\partial^{2}\zeta}^{\text{I}} & =-\frac{2(1+C_{2}\bar{N})}{C_{2}^{2}H^{2}\bar{N}^{2}a^{2}}\bigg((h_{2}^{\prime\prime}+2h_{2}^{\prime})(1+C_{2}\bar{N})+H(J_{3}^{\prime\prime}+2C_{2}J_{3}^{\prime}\bar{N}+C_{2}J_{3}^{\prime\prime}\bar{N})\bigg),\\
	\hat{\mathcal{C}}_{\partial_{i}\dot{\zeta}\partial^{i}\dot{\zeta}\partial^{2}\zeta}^{\text{I}} & =\frac{c_{4}^{(0;3,0)}(1+C_{2}\bar{N})^{2}}{2C_{2}^{3}H^{3}\bar{N}^{3}a^{4}(\omega_{2}+3H\omega_{3})}\bigg[2(h_{2}+h_{2}^{\prime})(1+C_{2}\bar{N})\nonumber \\
	& \ \ \ \ \ \ \ \ \ \ \ \ \ \ \ \ \ \ \ \ +H\bigg(2J_{3}^{\prime}(1+C_{2}\bar{N})+C_{2}(c_{1}^{(1;1,0)}+2J_{3})\bar{N}\bigg)\bigg],\\
	\hat{\mathcal{C}}_{\partial_{i}\dot{\zeta}\partial_{j}\dot{\zeta}\partial^{i}\partial^{j}\zeta}^{\text{I}} & =-\frac{3c_{4}^{(0;3,0)}(1+C_{2}\bar{N})^{2}}{2C_{2}^{3}H^{3}\bar{N}^{3}a^{4}(\omega_{2}+3H\omega_{3})}\bigg[2(h_{2}+2h_{2}^{\prime})(1+C_{2}\bar{N})\nonumber \\
	& \ \ \ \ \ \ \ \ \ \ \ \ \ \ \ \ \ \ \ \ +H\bigg(2J_{3}^{\prime}(1+C_{2}\bar{N})+C_{2}(c_{1}^{(1;1,0)}+2J_{3})\bar{N}\bigg)\bigg],
\end{align}
where the symbol ``$\simeq$'' denotes equality up to integrations by parts with respect to spatial derivatives. Varying with respect to $\zeta$ shows that each of these three terms generates an independent second-order time-derivative contribution to the equation of motion for $\zeta$. Hence, all three coefficients must vanish separately, i.e.,
\begin{align}
	\hat{\mathcal{C}}_{\dot{\zeta}^{2}\partial^{2}\zeta} & =\hat{\mathcal{C}}_{\partial_{i}\dot{\zeta}\partial^{i}\dot{\zeta}\partial^{2}\zeta}=\hat{\mathcal{C}}_{\partial_{i}\dot{\zeta}\partial_{j}\dot{\zeta}\partial^{i}\partial^{j}\zeta}=0. \label{cond_caseI}
\end{align}

Recall from (\ref{caseI_b3t}) and (\ref{b3tld}) that Case I requires $\tilde{b}_{3}\equiv c_{2}^{(0;3,0)}+3c_{4}^{(0;3,0)}=0$.
Depending on whether $c_{2}^{(0;3,0)}$ and $c_{4}^{(0;3,0)}$ vanish identically, the solutions to (\ref{cond_caseI}) split into two subcases.

The first subcase, which we call Case I.1, is
\begin{equation}
	c_{2}^{(0;3,0)}=c_{4}^{(0;3,0)}=0.
\end{equation}
Since $\tilde{b}_{3}\equiv c_{2}^{(0;3,0)}+3c_{4}^{(0;3,0)}=0$, the requirement $c_{4}^{(0;3,0)}=0$, which sets
$\hat{\mathcal{C}}_{\partial_{i}\dot{\zeta}\partial^{i}\dot{\zeta}\partial^{2}\zeta}=0$
and $\hat{\mathcal{C}}_{\partial_{i}\dot{\zeta}\partial_{j}\dot{\zeta}\partial^{i}\partial^{j}\zeta}=0$, also implies $c_{2}^{(0;3,0)}=0$. 
Imposing in addition $\hat{\mathcal{C}}_{\dot{\zeta}^{2}\partial^{2}\zeta} = 0$ leads to
\begin{align}
	h_{2}^{\prime\prime}+2h_{2}^{\prime} & =0,\label{eq:cubic constraint I,1 1}\\
	J_{3}^{\prime\prime}+2C_{2}J_{3}^{\prime}\bar{N}+C_{2}J_{3}^{\prime\prime}\bar{N} & =0.\label{eq:cubic constraint I,1 2}
\end{align}
Solving these equations for $h_{2}$ and $J_{3}$ gives
\begin{align}
	h_{2} & =\frac{C_{3}}{N}+C_{4},\\
	J_{3} & =D_{3}+\frac{D_{4}}{C_{2}(1+C_{2}N)}.
\end{align}

The second subcase, which we call Case I.2, is
\begin{equation}
	c_{2}^{(0;3,0)}=-3c_{4}^{(0;3,0)}\neq0.
\end{equation}
In this case, the conditions (\ref{cond_caseI}) yield
\begin{align}
	h_{2}^{\prime\prime}+2h_{2}^{\prime} & =0,\label{eq:cubic constraint I,2 1}\\
	J_{3}^{\prime\prime}+2C_{2}J_{3}^{\prime}\bar{N}+C_{2}J_{3}^{\prime\prime}\bar{N} & =0,\label{eq:cubic constraint I,2 2}\\
	h_{2}+h_{2}^{\prime} & =0,\label{eq:cubic constraint I,2 3}\\
	2J_{3}^{\prime}(1+C_{2}\bar{N})+C_{2}(c_{1}^{(1;1,0)}+2J_{3})\bar{N} & =0.\label{eq:cubic constraint I,2 4}
\end{align}
Equation \eqref{eq:cubic constraint I,2 3} imposes a stronger constraint on $h_{2}$ than \eqref{eq:cubic constraint I,2 1}. We can therefore solve $h_{2}$ and $J_{3}$ as
\begin{align}
	h_{2} & =\frac{C_{3}}{N},\\
	J_{3} & =D_{3}+\frac{D_{4}}{C_{2}(1+C_{2}N)}.
\end{align}
Substituting $J_{3}$ into \eqref{eq:cubic constraint I,2 4} then gives the solution for $c_{1}^{(1;1,0)}$,
\begin{align}
	c_{1}^{(1;1,0)} & =-2J_{3}^{\prime}\bigg(\frac{1+C_{2}N}{C_{2}}\bigg)-2J_{3}=-2D_{3}.
\end{align}

\begin{itemize}
	\item{\textbf{Terms with a single time derivative}}
\end{itemize}

We next examine the terms containing a single time derivative. They take the form
\begin{align}
	\mathcal{L}_{3,t1}^{\text{I}} & \simeq\hat{\mathcal{C}}_{\partial_{k}\dot{\zeta}\partial^{k}\partial^{2}\zeta\partial^{2}\zeta}^{\text{I}}\partial_{k}\dot{\zeta}\partial^{k}\partial^{2}\zeta\partial^{2}\zeta+\hat{\mathcal{C}}_{\partial^{i}\dot{\zeta}\partial_{i}\partial_{j}\zeta\partial^{j}\partial^{2}\zeta}^{\text{I}}\partial^{i}\dot{\zeta}\partial_{i}\partial_{j}\zeta\partial^{j}\partial^{2}\zeta\nonumber \\
	& \ \ \ \ +\hat{\mathcal{C}}_{\dot{\zeta}\partial^{i}\partial^{j}\zeta\partial_{i}\partial_{j}\zeta}^{\text{I}}\dot{\zeta}\partial^{i}\partial^{j}\zeta\partial_{i}\partial_{j}\zeta+\hat{\mathcal{C}}_{\dot{\zeta}\partial^{2}\zeta\partial^{2}\zeta}^{\text{I}}\dot{\zeta}\partial^{2}\zeta\partial^{2}\zeta\nonumber \\
	& \ \ \ \ +\hat{\mathcal{C}}_{\dot{\zeta}\zeta\partial^{2}\zeta}^{\text{I}}\dot{\zeta}\zeta\partial^{2}\zeta+\hat{\mathcal{C}}_{\dot{\zeta}\partial^{i}\zeta\partial_{i}\zeta}^{\text{I}}\dot{\zeta}\partial^{i}\zeta\partial_{i}\zeta+\hat{\mathcal{C}}_{\partial^{i}\dot{\zeta}\zeta\partial_{i}\zeta}^{\text{I}}\partial^{i}\dot{\zeta}\zeta\partial_{i}\zeta\nonumber+\hat{\mathcal{C}}_{\dot{\zeta}\zeta^{2}}^{\text{I}}\dot{\zeta}\zeta^{2}.
\end{align}
Since cancellations can occur only among terms with the same number of spatial derivatives, we discuss the different sectors separately.

\begin{enumerate}
	\item \textbf{Terms with no spatial-derivative:}\\
	The only term without spatial derivatives is $\hat{\mathcal{C}}_{\dot{\zeta}\zeta^{2}}\dot{\zeta}\zeta^{2}$. It does not contribute to the dynamics of $\zeta$, because it can be reduced by integration by parts:
	\begin{equation}
		\hat{\mathcal{C}}_{\dot{\zeta}\zeta^{2}}\dot{\zeta}\zeta^{2}\simeq-\frac{1}{3\bar{N}a^{3}}\partial_{t}(a^{3}\hat{\mathcal{C}}_{\dot{\zeta}\zeta^{2}})\zeta^{3}.
	\end{equation}
	
	\item \textbf{Terms with two spatial derivatives:}\\
	Among the terms involving two spatial derivatives, two are not independent because
	\begin{align}
		\hat{\mathcal{C}}_{\dot{\zeta}\zeta\partial^{2}\zeta}\dot{\zeta}\zeta\partial^{2}\zeta & \simeq-\frac{1}{2}\hat{\mathcal{C}}_{\dot{\zeta}\zeta\partial^{2}\zeta}\dot{\zeta}\partial^{i}\zeta\partial_{i}\zeta+\frac{1}{2\bar{N}a^{3}}\partial_{t}(a^{3}\hat{\mathcal{C}}_{\dot{\zeta}\zeta\partial^{2}\zeta})\zeta\partial_{i}\zeta\partial^{i}\zeta;\\
		\hat{\mathcal{C}}_{\partial^{i}\dot{\zeta}\zeta\partial_{i}\zeta}\partial^{i}\dot{\zeta}\zeta\partial_{i}\zeta & \simeq-\frac{1}{2}\hat{\mathcal{C}}_{\partial^{i}\dot{\zeta}\zeta\partial_{i}\zeta}\dot{\zeta}\partial_{i}\zeta\partial^{i}\zeta-\frac{1}{2\bar{N}a^{3}}\partial_{t}(a^{3}\hat{\mathcal{C}}_{\partial^{i}\dot{\zeta}\zeta\partial_{i}\zeta})\zeta\partial_{i}\zeta\partial^{i}\zeta.
	\end{align}
	The contribution from these terms vanishes identically in the cubic Lagrangian, because
	\begin{align}
		\hat{\mathcal{C}}_{\dot{\zeta}\partial^{i}\zeta\partial_{i}\zeta}^{\text{I}}-\frac{1}{2}\hat{\mathcal{C}}_{\dot{\zeta}\zeta\partial^{2}\zeta}^{\text{I}}-\frac{1}{2}\hat{\mathcal{C}}_{\partial^{i}\dot{\zeta}\zeta\partial_{i}\zeta}^{\text{I}} & =0.
	\end{align}
	Therefore, no additional constraint arises from the terms with two spatial derivatives.
	
	\item \textbf{Terms with six spatial derivatives:}\\
	For terms containing six spatial derivatives, one may use the integration-by-parts relation
	\begin{align}
		\hat{\mathcal{C}}_{\partial_{k}\dot{\zeta}\partial^{k}\partial^{2}\zeta\partial^{2}\zeta}\partial_{k}\dot{\zeta}\partial^{k}\partial^{2}\zeta\partial^{2}\zeta & \simeq\frac{1}{6\bar{N}a^{3}}\partial_{t}\bigg(a^{3}\hat{\mathcal{C}}_{\partial_{k}\dot{\zeta}\partial^{k}\partial^{2}\zeta\partial^{2}\zeta}\bigg)\partial^{2}\zeta\partial^{2}\zeta\partial^{2}\zeta.
	\end{align}
	Hence this term does not contribute to the dynamics of $\zeta$, and it may be neglected. Moreover, the coefficient $\hat{\mathcal{C}}{\partial^{i}\dot{\zeta}\partial{i}\partial_{j}\zeta\partial^{j}\partial^{2}\zeta}$ vanishes identically for all solutions obtained in Case I. Therefore, no additional constraint is induced from the terms with six spatial derivatives.
	
	\item \textbf{Terms with four spatial derivatives:}\\
	For the terms involving four spatial derivatives, the two structures are not independent, since
	\begin{align}
		\hat{\mathcal{C}}_{\dot{\zeta}\partial^{i}\partial^{j}\zeta\partial_{i}\partial_{j}\zeta}\dot{\zeta}\partial^{i}\partial^{j}\zeta\partial_{i}\partial_{j}\zeta & \simeq\hat{\mathcal{C}}_{\dot{\zeta}\partial^{i}\partial^{j}\zeta\partial_{i}\partial_{j}\zeta}\dot{\zeta}\partial^{2}\zeta\partial^{2}\zeta.
	\end{align}
	Accordingly, only one condition has to be imposed, namely $\hat{\mathcal{C}}_{\dot{\zeta}\partial^{i}\partial^{j}\zeta\partial_{i}\partial_{j}\zeta}^{\text{I}}+\hat{\mathcal{C}}_{\dot{\zeta}\partial^{2}\zeta\partial^{2}\zeta}^{\text{I}}=0$, where
	\begin{align}
		\hat{\mathcal{C}}_{\dot{\zeta}\partial^{i}\partial^{j}\zeta\partial_{i}\partial_{j}\zeta}^{\text{I}}+\hat{\mathcal{C}}_{\dot{\zeta}\partial^{2}\zeta\partial^{2}\zeta}^{\text{I}} & =\frac{H^{4}\Delta_{\dot{\zeta}\partial^{2}\zeta\partial^{2}\zeta,4}^{\text{I}}+H^{3}\Delta_{\dot{\zeta}\partial^{2}\zeta\partial^{2}\zeta,3}^{\text{I}}+H^{2}\Delta_{\dot{\zeta}\partial^{2}\zeta\partial^{2}\zeta,2}^{\text{I}}+H\Delta_{\dot{\zeta}\partial^{2}\zeta\partial^{2}\zeta,1}^{\text{I}}+\Delta_{\dot{\zeta}\partial^{2}\zeta\partial^{2}\zeta,0}^{\text{I}}}{6C_{2}^{3}H^{3}\bar{N}^{4}(C_{1}+C_{1}C_{2}\bar{N}+3D_{1}H\bar{N})a^{3}(\omega_{2}+3H\omega_{3})^{2}}. 
	\end{align}
	This relation must hold independently of the value of the Hubble parameter $H$, which implies that the coefficients of different powers of $H$ must vanish separately. We therefore obtain five equations,
	\begin{equation}         \Delta_{\dot{\zeta}\partial^{2}\zeta\partial^{2}\zeta,i}^{\text{I}}=0, \ \  i=0,...,4,\label{eq:DeltaI}
	\end{equation}
	whose explicit expressions are given in Appendix \ref{sec:Explicit form of the coefficients in case I}.
	From
	\begin{align}
		\Delta_{\dot{\zeta}\partial^{2}\zeta\partial^{2}\zeta,0}^{\text{I}} & =-4C_{4}^{2}\bar{N}(1+C_{2}\bar{N})^{4}\bigg(2C_{2}\omega_{2}^{2}+C_{1}(\omega_{2}-\omega_{2}^{\prime})\bigg) =0.
	\end{align}
	we may obtain either $C_{4}=0$ or $\omega_{2}=\frac{C_{1}N}{C_{5}-2C_{2}N}$.
	Accordingly, we further divide Case I.1 into two branches: the branch with $\omega_{2} = \frac{C_{1}N}{C_{5}-2C_{2}N}$ and $C_{4}\neq 0$, which we call Case I.11, and the branch with $C_{4}=0$, which we call Case I.12.
	
\end{enumerate}

\subsubsection{Case I.11}

We first consider Case I.11, which corresponds to
\begin{equation}
	\omega_{2}=\frac{C_{1}N}{C_{5}-2C_{2}N}\quad \text{and}\quad C_{4}\neq 0.
\end{equation}
The equation $\Delta_{\dot{\zeta}\partial^{2}\zeta\partial^{2}\zeta,1}^{\text{I11}}=0$, with
\begin{align}    \Delta_{\dot{\zeta}\partial^{2}\zeta\partial^{2}\zeta,1}^{\text{I11}}&=\frac{1}{(C_{5}-2C_{2}\bar{N})^{2}}4C_{1}C_{4}H\bar{N}(1+C_{2}\bar{N})^{3}\Big[C_{1}C_{2}\bar{N}^{2}\big(-C_{5}c_{1}^{(1;1,0)\prime}+2C_{2}(2D_{3}\nonumber \\
	&\ \ \ \ +c_{1}^{(1;1,0)}+c_{1}^{(1;1,0)\prime})\bar{N}\big)+3C_{4}\big(2C_{2}D_{1}\bar{N}^{3}-C_{2}(C_{5}-4)C_{5}\bar{N}(\omega_{3}-\omega_{3}^{\prime})\nonumber \\
	&\ \ \ \ -4C_{2}^{2}(C_{5}-1)\bar{N}^{2}\omega_{3}^{\prime}+C_{5}^{2}(\omega_{3}^{\prime}-2\omega_{3})+4C_{2}^{3}\bar{N}^{3}(\omega_{3}^{\prime}+\omega_{3})\big)\Big],
\end{align}
suggests the following solution for $\omega_3$:
\begin{align}
	\omega_{3}&=-\frac{\Big[-3C_{4}(C_{6}-2C_{2}D_{1}\bar{N})+C_{1}C_{2}\big(-C_{5}c_{1}^{(1;1,0)}+2C_{2}(2D_{3}+c_{1}^{(1;1,0)})\bar{N}\big)\Big]\bar{N}^{2}}{3C_{4}(C_{5}-2C_{2}\bar{N})^{2}(1+C_{2}\bar{N})}.
\end{align}

After substituting the solutions for $\omega_{2}$ and $\omega_{3}$, the remaining three equations can be recast into the matrix form
\begin{equation}\label{eq:MC=D}
	\bm{M}\cdot\bm{C}=\bm{D}
\end{equation}
with
\begin{equation}   
	\bm{M}=\begin{pmatrix}\hat{\mathcal{C}}{}_{4}^{3} & \hat{\mathcal{C}}{}_{4}^{2} & \hat{\mathcal{C}}{}_{4}^{1} & \hat{\mathcal{C}}{}_{4}^{0}\\
		\hat{\mathcal{C}}{}_{3}^{3} & \hat{\mathcal{C}}{}_{3}^{2} & \hat{\mathcal{C}}{}_{3}^{1} & \hat{\mathcal{C}}{}_{3}^{0}\\
		\hat{\mathcal{C}}{}_{2}^{3} & \hat{\mathcal{C}}{}_{2}^{2} & \hat{\mathcal{C}}{}_{2}^{1} & \hat{\mathcal{C}}{}_{2}^{0}
	\end{pmatrix},
\end{equation}  
\begin{align}      \bm{C}&=\begin{bmatrix}\big(c_{1}^{(1;1,0)}\big)^{3},&\big(c_{1}^{(1;1,0)}\big)^{2},&\big(c_{1}^{(1;1,0)}\big)^{1},&\big(c_{1}^{(1;1,0)}\big)^{0}\end{bmatrix}^{\text{T}},
\end{align}
and
\begin{align}      
	\bm{D}&=\begin{pmatrix}M_{4}c_{1}^{(1;1,0)\prime}, &M_{3}c_{1}^{(1;1,0)\prime}, &M_{2}c_{1}^{(1;1,0)\prime}\end{pmatrix}^{\text{T}},
\end{align}
where $M_{i}$ and $\hat{\mathcal{C}}{}_{i}^{j}$ are particular functions of the lapse corresponding to the $i$-th equation. 
To solve the algebraic system \eqref{eq:MC=D}, we must distinguish between the homogeneous and non-homogeneous cases, depending on the degeneracy structure of $\bm{D}$.

Note that the condition $M_{2}=0$ is equivalent to 
\begin{equation} \label{eq:M2=0}  c_{1}^{(1;1,0)}=\frac{2(C_{1}C_{2}C_{5}D_{3}-3C_{4}D_{6}+6C_{2}C_{4}D_{1}\bar{N}+2C_{1}C_{2}^{2}D_{3}\bar{N})}{C_{1}C_{2}(C_{5}-2C_{2}\bar{N})}.
\end{equation}
The condition $M_{3}=0$ implies 
\begin{equation}    
	c_{1}^{(1;1,0)}=-\frac{6C_{4}D_{1}}{C_{1}C_{2}}-2D_{3},
\end{equation}
or
\begin{equation}\label{eq:M3=0,b}    c_{1}^{(1;1,0)}=\frac{2(C_{1}C_{2}C_{5}D_{3}-3C_{4}D_{6}+6C_{2}C_{4}D_{1}\bar{N}+2C_{1}C_{2}^{2}D_{3}\bar{N})}{C_{1}C_{2}(C_{5}-2C_{2}\bar{N})}.
\end{equation}
The condition $M_{4}=0$ means
\begin{equation}\label{eq:M4=0,a}
	D_{1}=0,
\end{equation}   
or
\begin{equation}
	c_{1}^{(1;1,0)}=-2D_{3},
\end{equation}   
or
\begin{equation}\label{eq:M4=0,c}    c_{1}^{(1;1,0)}=\frac{2(C_{1}C_{2}C_{5}D_{3}-3C_{4}D_{6}+6C_{2}C_{4}D_{1}\bar{N}+2C_{1}C_{2}^{2}D_{3}\bar{N})}{C_{1}C_{2}(C_{5}-2C_{2}\bar{N})}.
\end{equation}
The intersection of these three conditions, namely $\bm{D}=0$, requires \eqref{eq:M2=0}, \eqref{eq:M3=0,b}, and \eqref{eq:M4=0,c} to hold simultaneously. 

We first consider the non-homogenous case in which all components of the matrix $\bm{D}$ are nonzero. The consistency of the equations then immediately implies 
\begin{align}    
	M_{3}M_{2}\hat{\mathcal{C}}{}_{4}^{3}=M_{4}M_{2}\hat{\mathcal{C}}{}_{3}^{3}&=M_{4}M_{3}\hat{\mathcal{C}}{}_{2}^{3},\\
	M_{3}M_{2}\hat{\mathcal{C}}{}_{4}^{2}=M_{4}M_{2}\hat{\mathcal{C}}{}_{3}^{2}&=M_{4}M_{3}\hat{\mathcal{C}}{}_{2}^{2},\\
	M_{3}M_{2}\hat{\mathcal{C}}{}_{4}^{1}=M_{4}M_{2}\hat{\mathcal{C}}{}_{3}^{1}&=M_{4}M_{3}\hat{\mathcal{C}}{}_{2}^{1},\\
	M_{3}M_{2}\hat{\mathcal{C}}{}_{4}^{1}=M_{4}M_{2}\hat{\mathcal{C}}{}_{3}^{1}&=M_{4}M_{3}\hat{\mathcal{C}}{}_{2}^{1}.
\end{align}
due to the uniqueness of $c_{1}^{(1;1,0)\prime}$. However, these relations are incompatible with the non-homogeneous conditions $M_{2}\neq 0$, $M_{3}\neq 0$, and $M_{4}\neq 0$. Hence, this case admits no solution.

We now turn to the cases in which some entries of $\bm{D}$ vanish. 

\begin{itemize}
	\item $M_2=M_3=M_4=0$
\end{itemize}
In this case, besides the necessary condition \eqref{eq:M2=0}, one must further impose $D_{1}=0$ and $D_{4}=0$ in order to satisfy the equations $\Delta_{\dot{\zeta}\partial^{2}\zeta\partial^{2}\zeta,i}^{\text{I}} =0$ with $i=2,3,4$. 

If $c_{1}^{(1;1,0)} = c_{1}^{(1;1,0)}(t,N)$, we obtain the following set of solutions for the coefficients:
\begin{align}   
	C_{4} \neq0,\quad \omega_{2}=\frac{C_{1}N}{C_{5}-2C_{2}N}, \quad \omega_{3}&=\frac{(2C_{1}C_{2}C_{5}D_{3}-3C_{4}C_{6})N^{2}}{3C_{4}(1+C_{2}N)(C_{5}-2C_{2}N)^{2}},
\end{align}
and
\begin{align}    b_{3}=0,\quad J_{3}=D_{3}, \quad c_{1}^{(1;1,0)}=\frac{-6C_{4}C_{6}+2C_{1}C_{2}D_{3}(C_{5}+2C_{2}N)}{C_{1}C_{2}(C_{5}-2C_{2}N)}.
\end{align}
We refer to this set as ``solution A1''.
The corresponding action takes the form
\begin{align}
	S^{\text{A1}}	&=\int\text{d}t\text{d}^{3}x\!N\sqrt{h}\Big[\frac{C_{1}N}{C_{5}-2C_{2}N}\hat{K}^{ij}\hat{K}_{ij}+\frac{1}{3}\frac{C_{1}N}{1+C_{2}N}K^{2}+\Big(\frac{C_{3}}{N}+C_{4}\Big){}^{3}\!R+\Big(\frac{B_{1}}{N}+B_{2}\Big)\nonumber \\
	&\ \ \ \ +c_{1}^{(0;3,0)}\hat{K}_{ij}\hat{K}^{jk}\hat{K}_{k}^{i}+\frac{(2C_{1}C_{2}C_{5}D_{3}-3C_{4}C_{6})N^{2}}{3C_{4}(1+C_{2}N)(C_{5}-2C_{2}N)^{2}}\hat{K}_{ij}\hat{K}^{ij}K\nonumber \\
	&\ \ \ \ +\Big(\frac{2C_{1}C_{2}D_{3}(C_{5}+2C_{2}N)-6C_{4}C_{6}}{C_{1}C_{2}(C_{5}-2C_{2}N)}\Big){}^{3}\!R^{ij}K_{ij}+\frac{1}{3}D_{3}{}^{3}\!RK\Big], \label{S_A1_pre}
\end{align}
with $C_{4}\neq0$.
Note that in (\ref{S_A1_pre}), and likewise below, we rewrite the action in a more compact form by expressing the extrinsic curvature as $K_{ij}=\hat{K}_{ij}+\frac{1}{3}h_{ij}K$, where $\hat{K}_{ij}$ and $K$ denote the traceless part and the trace of $K_{ij}$, respectively. 

\begin{itemize}
	\item $M_{2}\neq0$ and $M_{3}=M_{4}=0$
\end{itemize}

The conditions $M_{3}=M_{4}=0$ imply $c_{1}^{(1;1,0)}=-2D_{3}$ and $ D_{1}=0$. The equations $\Delta_{\dot{\zeta}\partial^{2}\zeta\partial^{2}\zeta,i}^{\text{I11}}=0$ with $i=2,3,4$ imply $D_{4}=0$ and $C_{6}=\frac{2}{3C_{4}}C_{1}C_{2}C_{5}D_{3}$, which in turn gives $\omega_{3}=0$. However, substituting $c_{1}^{(1;1,0)}=-2D_{3}$ and $D_{1}=0$ into the condition $M_{2}\neq0$ immediately yields $C_{6}\neq\frac{2}{3C_{4}}C_{1}C_{2}C_{5}D_{3}$, in obvious contradiction with the above result.

\begin{itemize}
	\item $M_{2}\neq0$, $M_{3}\neq0$ and $M_{4}=0$
\end{itemize}

One solution extracted from $M_{3}\neq0$ and $M_{4}=0$ is $c_{1}^{(1;1,0)}=-2D_{3}$ and $D_{1}\neq0$, while the condition $M_{2}\neq0$ requires $C_{6}\neq\frac{2}{3C_{4}}C_{1}C_{2}C_{5}D_{3}$. The equations $\Delta_{\dot{\zeta}\partial^{2}\zeta\partial^{2}\zeta,i}^{\text{I11}}=0$ with $i=2,3,4$ also imply $D_{4}=0$, $=\frac{1}{3C_{4}}(2C_{1}C_{2}C_{5}D_{3}+3C_{4}C_{5}D_{1})$. This branch can be consistent with $M_{2}\neq0$.

We thus obtain the set of solutions
\begin{align}
	C_{4}\neq0,	\quad \omega_{2} =\frac{C_{1}N}{C_{5}-2C_{2}N}, \quad C_{5}\neq0, \quad \omega_{3}	=\frac{D_{1}N^{2}}{(1+C_{2}N)(C_{5}-2C_{2}N)}\neq0,
\end{align}
and
\begin{align}    
	b_{3}\neq0, \quad J_{3}=D_{3},\quad c_{1}^{(1;1,0)}=-2D_{3}.
\end{align}
We refer to this set as ``solution A2''.
The corresponding action is
\begin{align}
	S^{\text{A2}}	&=\int\text{d}t\text{d}^{3}x\!N\sqrt{h}\Big[\frac{C_{1}N}{C_{5}-2C_{2}N}\hat{K}^{ij}\hat{K}_{ij}+\frac{1}{3}\frac{C_{1}N}{1+C_{2}N}K^{2}+\Big(\frac{C_{3}}{N}+C_{4}\Big){}^{3}\!R+\Big(\frac{B_{1}}{N}+B_{2}\Big)\nonumber \\
	&\ \ \ \ +c_{1}^{(0;3,0)}\hat{K}_{ij}\hat{K}^{jk}\hat{K}_{k}^{i}+\frac{D_{1}N^{2}}{(1+C_{2}N)(C_{5}-2C_{2}N)}\hat{K}_{ij}\hat{K}^{ij}K+\frac{1}{9}\frac{D_{1}N^{2}}{(1+C_{2}N)^{2}}K^{3}\nonumber \\
	&\ \ \ \ -2D_{3}{}^{3}\!R^{ij}K_{ij}+\frac{1}{3}D_{3}{}^{3}\!RK\Big],
\end{align}
with $C_{4}\neq0$, $C_{5}\neq0$, $D_{1}\neq0$.

The remaining possibility is $D_{1}=0$, $c_{1}^{(1;1,0)}\neq-2D_{3}$, and $c_{1}^{(1;1,0)}\neq-2D_{3}+\frac{4D_{3}C_{5}C_{1}C_{2}-6C_{4}D_{6}}{C_{1}C_{2}(C_{5}-2C_{2}\bar{N})}$.
The equation $\Delta_{\dot{\zeta}\partial^{2}\zeta\partial^{2}\zeta,4}^{\text{I11}}=0$ implies either $c_{1}^{(1;1,0)}=\frac{4D_{3}C_{1}C_{2}^{2}N-3C_{4}C_{6}}{C_{1}C_{2}(C_{5}-2C_{2}N)}$ and $D_{4}\neq0$ or $D_{4}=0$. After substituting the former possibility into $\Delta_{\dot{\zeta}\partial^{2}\zeta\partial^{2}\zeta,3}^{\text{I11}}$, the equations with $i=3,4$ are satisfied automatically. One then only needs to consider $\Delta_{\dot{\zeta}\partial^{2}\zeta\partial^{2}\zeta,2}^{\text{I11}}=0$, which vanishes only if $D_{4}=0$. Hence this possibility is not viable. In the other case, namely $D_{4}=0$, the expression $\Delta_{\dot{\zeta}\partial^{2}\zeta\partial^{2}\zeta,4}^{\text{I11}}$ vanishes, while $\Delta_{\dot{\zeta}\partial^{2}\zeta\partial^{2}\zeta,3}^{\text{I11}}=\Delta_{\dot{\zeta}\partial^{2}\zeta\partial^{2}\zeta,2}^{\text{I11}}=0$ implies $c_{1}^{(1;1,0)}=\frac{-6C_{4}C_{6}+2D_{3}C_{1}C_{2}(C_{5}+2C_{2}\bar{N})}{C_{1}C_{2}(C_{5}-2C_{2}\bar{N})}$, which is again inconsistent.

\begin{itemize}
	\item $M_{2}\neq0$, $M_{3}=0$ and $M_{4}\neq0$
\end{itemize}

The conditions $M_{3}=0$ and $M_{4}\neq0$ imply $c_{1}^{(1;1,0)}=-\frac{6C_{4}D_{1}}{C_{1}C_{2}}-2D_{3}$
and $D_{1}\neq0$. The condition $M_{2}\neq0$ further requires 
$C_{6}\neq C_{5}D_{1}+\frac{2}{3C_{4}}C_{1}C_{2}C_{5}D_{3}$. Unfortunately, no solution can be obtained in this case.

\subsubsection{Case I.12}

We next turn to Case I.12, which corresponds to $C_{4}=0$.
For this branch, the coefficient $\Delta_{\dot{\zeta}\partial^{2}\zeta\partial^{2}\zeta,1}^{\text{I.12}}$ vanishes identically. We therefore begin with $\Delta_{\dot{\zeta}\partial^{2}\zeta\partial^{2}\zeta,2}^{\text{I.12}}$, which implies
\begin{align}
	\omega_{2}&=\frac{C_{1}C_{2}^{2}(2D_{3}+c_{1}^{(1;1,0)})^{2}N(1+C_{2}N)}{-8D_{4}^{2}+C_{1}C_{2}^{2}C_{5}(1+C_{2}N)}
\end{align}
and $c_{1}^{(1;1,0)}\neq-2D_{3}$. Substituting this expression for $\omega_{2}$ into $\Delta_{\dot{\zeta}\partial^{2}\zeta\partial^{2}\zeta,3}^{\text{I12}}$, we obtain the solution for $\omega_3$,
\begin{align}    \omega_{3}&=\frac{\big(-8C_{2}^{2}D_{1}D_{4}^{2}+C_{6}(1+C_{2}N)\big)(2D_{3}+c_{1}^{(1;1,0)})^{2}N^{2}}{\bigl(-8D_{4}^{2}+C_{1}C_{2}^{2}C_{5}(1+C_{2}N)\bigr)^{2}}
\end{align}
with $C_{5}\neq\frac{8D_{4}^{2}}{C_{1}C_{2}^{2}(1+C_{2}\bar{N})}$. Finally, the last equation $\Delta_{\dot{\zeta}\partial^{2}\zeta\partial^{2}\zeta,4}^{\text{I12}}=0$ gives two possible solutions:
\begin{align}
	D_{4}=0,	\qquad C_{5}\neq0, 
\end{align}
or
\begin{align}
	D_{4} \neq0, \qquad C_{6}=C_{1}C_{2}^{4}C_{5}D_{1}.
\end{align}

We therefore obtain two sets of solutions for the coefficients in Case I.12.
One set is
\begin{align}
	C_{4}=D_{4}=0,\qquad \omega_{2}=\frac{1}{C_{5}}(2D_{3}+c_{1}^{(1;1,0)})^{2}N,\qquad \omega_{3}=\frac{C_{6}(2D_{3}+c_{1}^{(1;1,0)})^{2}N^{2}}{C_{1}^{2}C_{2}^{4}C_{5}^{2}(1+C_{2}N)},
\end{align}
which we call ``Solution B1''.
The corresponding action can be written as
\begin{align}
	S^{\text{B1}}&=\int\text{d}t\text{d}^{3}x\!N\sqrt{h}\Big[\frac{1}{C_{5}}(2D_{3}+c_{1}^{(1;1,0)})^{2}N\hat{K}^{ij}\hat{K}_{ij}+\frac{1}{3}\frac{C_{1}N}{1+C_{2}N}K^{2}+\frac{C_{3}}{N}{}^{3}\!R+\Big(\frac{B_{1}}{N}+B_{2}\Big)\nonumber \\
	&\ \ \ \ +c_{1}^{(0;3,0)}\hat{K}_{ij}\hat{K}^{jk}\hat{K}_{k}^{i}+\frac{C_{6}(2D_{3}+c_{1}^{(1;1,0)})^{2}N^{2}}{C_{1}^{2}C_{2}^{4}C_{5}^{2}(1+C_{2}N)}\hat{K}_{ij}\hat{K}^{ij}K+\frac{1}{9}\frac{D_{1}N^{2}}{(1+C_{2}N)^{2}}K^{3}\nonumber \\
	&\ \ \ \ +c_{1}^{(1;1,0)}{}^{3}\!R^{ij}K_{ij}+\frac{1}{3}D_{3}{}^{3}\!RK\Big].
\end{align}

The other set of solutions is
\begin{align}
	C_{4}=0,\qquad \omega_{2} =\frac{C_{1}C_{2}^{2}(2D_{3}+c_{1}^{(1;1,0)})^{2}N(1+C_{2}N)}{-8D_{4}^{2}+C_{1}C_{2}^{2}C_{5}(1+C_{2}N)},\qquad \omega_{3}=\frac{C_{2}^{2}D_{1}(2D_{3}+c_{1}^{(1;1,0)})^{2}N^{2}}{-8D_{4}^{2}+C_{1}C_{2}^{2}C_{5}(1+C_{2}N)},
\end{align}
which we call ``Solution B2''.
The corresponding action is
\begin{align}
	S^{\text{B2}}	&=\int\text{d}t\text{d}^{3}x\!N\sqrt{h}\Big[\frac{C_{1}C_{2}^{2}(2D_{3}+c_{1}^{(1;1,0)})^{2}N(1+C_{2}N)}{-8D_{4}^{2}+C_{1}C_{2}^{2}C_{5}(1+C_{2}N)}\hat{K}^{ij}\hat{K}_{ij}+\frac{1}{3}\frac{C_{1}N}{1+C_{2}N}K^{2}+\frac{C_{3}}{N}{}^{3}\!R+\Big(\frac{B_{1}}{N}+B_{2}\Big)\nonumber \\
	&\ \ \ \ +c_{1}^{(0;3,0)}\hat{K}_{ij}\hat{K}^{jk}\hat{K}_{k}^{i}+\frac{C_{2}^{2}D_{1}(2D_{3}+c_{1}^{(1;1,0)})^{2}N^{2}}{-8D_{4}^{2}+C_{1}C_{2}^{2}C_{5}(1+C_{2}N)}\hat{K}_{ij}\hat{K}^{ij}K+\frac{1}{9}\frac{D_{1}N^{2}}{(1+C_{2}N)^{2}}K^{3}\nonumber \\
	&\ \ \ \ +c_{1}^{(1;1,0)}{}^{3}\!R^{ij}K_{ij}+\frac{1}{3}\Big(D_{3}+\frac{D_{4}}{C_{2}(1+C_{2}N)}\Big){}^{3}\!RK\Big]
\end{align}
with $D_{4}\neq0$.

\subsubsection{Case I.2}

Finally, we return to Case I.2, with $C_4=0$. 
In this branch, the equations \eqref{eq:DeltaI} are drastically simplified, which leads to $D_4=0$. We therefore obtain another set of solutions for the coefficients, which we call ``Solution C''. 
The corresponding action is
\begin{align}    S^{\text{C}}&=\int\text{d}t\text{d}^{3}xN\sqrt{h}\Big[\omega_{2}\hat{K}^{ij}\hat{K}_{ij}+\frac{1}{3}\frac{C_{1}N}{1+C_{2}N}K^{2}+\frac{C_{3}}{N}{}^{3}\!R+\Big(\frac{B_{1}}{N}+B_{2}\Big)\nonumber \\
	&\ \ \ \ +c_{1}^{(0;3,0)}\hat{K}_{ij}\hat{K}^{jk}\hat{K}_{k}^{i}+c_{2}^{(0;3,0)}\hat{K}_{ij}a^{i}a^{j}+\omega_{3}\hat{K}_{ij}\hat{K}^{ij}K+\frac{1}{9}\frac{D_{1}N^{2}}{(1+C_{2}N)^{2}}K^{3}\nonumber \\
	&\ \ \ \ -2D_{3}{}^{3}\!R^{ij}K_{ij}+\frac{1}{3}D_{3}{}^{3}\!RK\Big].
\end{align}

\subsection{Case II}
\label{subsec:3.2}

For case II, with the conditions (\ref{case2omg2}) and (\ref{eq:solution for omega3}), the action takes the form
\begin{align}
	S^{\text{II}} & =\int\text{d}t\text{d}^{3}xN\sqrt{h}\Big[-\frac{b_{2}(f_{3}-3\tilde{f}_{3})^{2}}{2f_{3}^{2}}K^{ij}K_{ij}+\frac{1}{3}\Big(\frac{C_{1}N}{1+C_{2}N}+\frac{b_{2}(f_{3}-3\tilde{f}_{3})^{2}}{2f_{3}^{2}}\Big)K^{2}\nonumber \\
	& \ \ \ \ +h_{2}{}^{3}\!R+\Big(\frac{B_{1}}{N}+B_{2}\Big)+\Big(-\frac{b_{3}(f_{3}-3\tilde{f}_{3})^{2}}{2f_{3}^{2}}-c_{3}^{(0;3,0)}\Big)K_{ij}K^{jk}K_{k}^{i}\nonumber \\
	& \ \ \ \ -3c_{4}^{(0;3,0)}K_{ij}a^{i}a^{j}+c_{3}^{(0;3,0)}K_{ij}K^{ij}K+c_{4}^{(0;3,0)}Ka^{i}a_{i}\nonumber \\
	& \ \ \ \ +\frac{1}{9}\Big(\frac{D_{1}N^{2}}{(1+C_{2}N)^{2}}-c_{1}^{(0;3,0)}-3c_{3}^{(0;3,0)}\Big)K^{3}+c_{1}^{(0;1,1)}K_{ij}\nabla^{i}a^{j}+c_{2}^{(0;1,1)}K\nabla_{i}a^{i}\nonumber \\
	& \ \ \ \ +c_{1}^{(1;1,0)}R^{ij}K_{ij}+\frac{1}{3}(J_{3}-c_{1}^{(1;1,0)})RK\Big].
\end{align}
The linear-order solutions for the auxiliary variables $A$ and $B$ are
\begin{align}
	A & =\frac{1}{\Delta_{1}}2(f_{3}-3\tilde{f}_{3})\bar{N}(1+C_{2}\bar{N})(C_{1}+C_{1}C_{2}\bar{N}+3D_{1}H\bar{N})a^{2}\dot{\zeta}\nonumber \\
	& \ \ \ \ +\frac{1}{3\Delta_{1}^{2}}\bigg[2(1+C_{2}\bar{N})^{3}\Bigl(f_{3}^{2}\tilde{f}_{3}(c_{1}^{(1;1,0)}+2J_{3})(1+C_{2}\bar{N})^{3}\partial^{4}\nonumber \\
	& \ \ \ \ \ \ \ \ -2\bar{N}(C_{1}+C_{1}C_{2}\bar{N}+3D_{1}H\bar{N})\bigl(C_{2}f_{3}^{2}H(c_{1}^{(1;1,0)}+2J_{3})\bar{N}\nonumber \\
	& \ \ \ \ \ \ \ \ +12f_{3}\tilde{f}_{3}(h_{2}+h_{2}^{\prime}+HJ_{3}^{\prime})(1+C_{2}\bar{N})-18\tilde{f}_{3}^{2}(h_{2}+h_{2}^{\prime}+HJ_{3}^{\prime})(1+C_{2}\bar{N})\bigr)a^{2}\partial^{2}\Bigr)\zeta\bigg], \label{sol_caseII_A}\\
	B & =\frac{1}{\Delta_{1}}f_{3}^{2}(1+C_{2}\bar{N})^{3}a\dot{\zeta}\nonumber \\
	& \ \ \ \ -\frac{1}{3\Delta_{1}^{2}}4(f_{3}+C_{2}f_{3}\bar{N})^{2}a\bigg[\bigg((1+C_{2}\bar{N})^{3}\bigl(C_{2}f_{3}H(c_{1}^{(1;1,0)}+2J_{3})\bar{N}\nonumber \\
	& \ \ \ \ \ \ \ \ +3\tilde{f}_{3}(h_{2}+h_{2}^{\prime}+HJ_{3}^{\prime})(1+C_{2}\bar{N})\bigr)\bigg)\partial^{2}-3C_{2}H\bar{N}^{2}(C_{1}+C_{1}C_{2}\bar{N}+3D_{1}H\bar{N})\times\nonumber \\
	& \ \ \ \ \ \ \ \ \bigl(C_{2}H(c_{1}^{(1;1,0)}+2J_{3})\bar{N}+2(h_{2}+h_{2}^{\prime}+HJ_{3}^{\prime})(1+C_{2}\bar{N})\bigr)a^{2}\Bigr)\bigg]\zeta, \label{sol_caseII_B}
\end{align}
with
\begin{align}
	\Delta_{1} & =f_{3}\tilde{f}_{3}(1+C_{2}\bar{N})^{3}\partial^{2}+2C_{2}(f_{3}-3\tilde{f}_{3})H\bar{N}^{2}(C_{1}+C_{1}C_{2}\bar{N}+3D_{1}H\bar{N})a^{2}.\label{eq:Delta1}
\end{align}

In Case II, the solutions for $A$ and $B$ generally involve nonlocal operators, which substantially complicates the analysis. The full derivation is presented in Appendix \ref{sec:Degeneracy Conditions: case II}.
After a careful analysis, we find that no 2-DOF theory can be constructed in this branch because the required constraints are mutually inconsistent. This result indicates that the cubic terms containing the acceleration, i.e., spatial derivatives of the lapse function, inevitably reintroduce an additional scalar degree of freedom beyond the two tensor modes.

\subsection{Summary of the explicit actions}
\label{subsec:3.3}

In summary, we obtain five actions:
\begin{align}
	S^{\text{A1}}	&=\int\text{d}t\text{d}^{3}x\!N\sqrt{h}\Big[\frac{C_{1}N}{C_{5}-2C_{2}N}\hat{K}^{ij}\hat{K}_{ij}+\frac{1}{3}\frac{C_{1}N}{1+C_{2}N}K^{2}+\Big(\frac{C_{3}}{N}+C_{4}\Big){}^{3}\!R+\Big(\frac{B_{1}}{N}+B_{2}\Big)\nonumber \\
	&\ \ \ \ +c_{1}^{(0;3,0)}\hat{K}_{ij}\hat{K}^{jk}\hat{K}_{k}^{i}+\frac{(2C_{1}C_{2}C_{5}D_{3}-3C_{4}C_{6})N^{2}}{3C_{4}(1+C_{2}N)(C_{5}-2C_{2}N)^{2}}\hat{K}_{ij}\hat{K}^{ij}K\nonumber \\
	&\ \ \ \ +\Big(\frac{2C_{1}C_{2}D_{3}(C_{5}+2C_{2}N)-6C_{4}C_{6}}{C_{1}C_{2}(C_{5}-2C_{2}N)}\Big){}^{3}\!R^{ij}K_{ij}+\frac{1}{3}D_{3}{}^{3}\!RK\Big]\label{eq:S_A1}
\end{align}
with $C_{4}\neq0$, and
\begin{align}
	S^{\text{A2}}	&=\int\text{d}t\text{d}^{3}x\!N\sqrt{h}\Big[\frac{C_{1}N}{C_{5}-2C_{2}N}\hat{K}^{ij}\hat{K}_{ij}+\frac{1}{3}\frac{C_{1}N}{1+C_{2}N}K^{2}+\Big(\frac{C_{3}}{N}+C_{4}\Big){}^{3}\!R+\Big(\frac{B_{1}}{N}+B_{2}\Big)\nonumber \\
	&\ \ \ \ +c_{1}^{(0;3,0)}\hat{K}_{ij}\hat{K}^{jk}\hat{K}_{k}^{i}+\frac{D_{1}N^{2}}{(1+C_{2}N)(C_{5}-2C_{2}N)}\hat{K}_{ij}\hat{K}^{ij}K+\frac{1}{9}\frac{D_{1}N^{2}}{(1+C_{2}N)^{2}}K^{3}\nonumber \\
	&\ \ \ \ -2D_{3}{}^{3}\!R^{ij}K_{ij}+\frac{1}{3}D_{3}{}^{3}\!RK\Big]\label{eq:S_A2}
\end{align}
with $C_{4}\neq0$, $C_{5}\neq0$, $D_{1}\neq0$,
\begin{align}
	S^{\text{B1}}&=\int\text{d}t\text{d}^{3}x\!N\sqrt{h}\Big[\frac{1}{C_{5}}(2D_{3}+c_{1}^{(1;1,0)})^{2}N\hat{K}^{ij}\hat{K}_{ij}+\frac{1}{3}\frac{C_{1}N}{1+C_{2}N}K^{2}+\frac{C_{3}}{N}{}^{3}\!R+\Big(\frac{B_{1}}{N}+B_{2}\Big)\nonumber \\
	&\ \ \ \ +c_{1}^{(0;3,0)}\hat{K}_{ij}\hat{K}^{jk}\hat{K}_{k}^{i}+\frac{C_{6}(2D_{3}+c_{1}^{(1;1,0)})^{2}N^{2}}{C_{1}^{2}C_{2}^{4}C_{5}^{2}(1+C_{2}N)}\hat{K}_{ij}\hat{K}^{ij}K+\frac{1}{9}\frac{D_{1}N^{2}}{(1+C_{2}N)^{2}}K^{3}\nonumber \\
	&\ \ \ \ +c_{1}^{(1;1,0)}{}^{3}\!R^{ij}K_{ij}+\frac{1}{3}D_{3}{}^{3}\!RK\Big],\label{eq:S_B1}
\end{align}
\begin{align}
	S^{\text{B2}}	&=\int\text{d}t\text{d}^{3}x\!N\sqrt{h}\Big[\frac{C_{1}C_{2}^{2}(2D_{3}+c_{1}^{(1;1,0)})^{2}N(1+C_{2}N)}{-8D_{4}^{2}+C_{1}C_{2}^{2}C_{5}(1+C_{2}N)}\hat{K}^{ij}\hat{K}_{ij}+\frac{1}{3}\frac{C_{1}N}{1+C_{2}N}K^{2}+\frac{C_{3}}{N}{}^{3}\!R+\Big(\frac{B_{1}}{N}+B_{2}\Big)\nonumber \\
	&\ \ \ \ +c_{1}^{(0;3,0)}\hat{K}_{ij}\hat{K}^{jk}\hat{K}_{k}^{i}+\frac{C_{2}^{2}D_{1}(2D_{3}+c_{1}^{(1;1,0)})^{2}N^{2}}{-8D_{4}^{2}+C_{1}C_{2}^{2}C_{5}(1+C_{2}N)}\hat{K}_{ij}\hat{K}^{ij}K+\frac{1}{9}\frac{D_{1}N^{2}}{(1+C_{2}N)^{2}}K^{3}\nonumber \\
	&\ \ \ \ +c_{1}^{(1;1,0)}{}^{3}\!R^{ij}K_{ij}+\frac{1}{3}\Big(D_{3}+\frac{D_{4}}{C_{2}(1+C_{2}N)}\Big){}^{3}\!RK\Big]\label{eq:S_B2}
\end{align}
with $D_{4}\neq0$, and
\begin{align}    S^{\text{C}}&=\int\text{d}t\text{d}^{3}xN\sqrt{h}\Big[\omega_{2}\hat{K}^{ij}\hat{K}_{ij}+\frac{1}{3}\frac{C_{1}N}{1+C_{2}N}K^{2}+\frac{C_{3}}{N}{}^{3}\!R+\Big(\frac{B_{1}}{N}+B_{2}\Big)\nonumber \\
	&\ \ \ \ +c_{1}^{(0;3,0)}\hat{K}_{ij}\hat{K}^{jk}\hat{K}_{k}^{i}+c_{2}^{(0;3,0)}\hat{K}_{ij}a^{i}a^{j}+\omega_{3}\hat{K}_{ij}\hat{K}^{ij}K+\frac{1}{9}\frac{D_{1}N^{2}}{(1+C_{2}N)^{2}}K^{3}\nonumber \\
	&\ \ \ \ -2D_{3}{}^{3}\!R^{ij}K_{ij}+\frac{1}{3}D_{3}{}^{3}\!RK\Big].\label{eq:S_C}
\end{align}
For all five actions, the scalar mode is completely eliminated up to cubic order in perturbations around a cosmological background. Equivalently, these five actions propagate 2 DOFs up to cubic order in perturbation theory. 

At this stage, however, it is important to note that although each of the above actions is free of scalar-mode propagation at cubic order, the quadratic part of the action, namely the terms quadratic in the extrinsic curvature, is incompatible with the Einstein--Hilbert action when $C_4=0$ ($C_3$ is independent of the lapse function). In this sense, $S^{\text{B1}}$, $S^{\text{B2}}$, and $S^{\text{C}}$ do not admit a GR limit. Taking this into account, we shall focus on the actions $S^{\text{A1}}$ and $S^{\text{A2}}$.

\section{Relations with Other Theories} \label{sec:4}

In this section, we discuss the relation between our theory and previously known 2-DOF theories in the literature, including the cuscuton theory \cite{Afshordi:2006ad} and its extension \cite{Iyonaga:2018vnu}, as well as the quadratic 2-DOF theory \cite{Hu:2021yaq}. We will show that these theories are compatible with our theory.

\subsection{Cuscuton and Extended Cuscuton Theories}
\label{subsec:4.1}

The cuscuton theory proposed in \cite{Afshordi:2006ad} is the first example of a gravity theory without propagating scalar degrees of freedom. As a special case of $k$-essence theory, its action can be expressed in the unitary gauge as
\begin{align}
	S_{\text{cus.}} & =\int\text{d}t\text{d}^{3}xN\sqrt{h}\Big(\hat{K}{}^{ij}\hat{K}_{ij}-\frac{2}{3}K^{2}+R+\frac{1}{N}\Big).
\end{align}
It is straightforward to see that this is a subclass of the action $S^{\mathrm{A1}}$ \eqref{eq:S_A1} obtained by taking
\begin{align}   c_{1}^{(0;3,0)}&=0,&B_{2}&=C_{3}=C_{6}=D_{3}=0,&B_{1}&=C_{4}=1,&C_{1}&=-2C_{2}\neq0, &C_{1},C_{2}&\gg1.
\end{align}

The extended cuscuton theory was developed in \cite{Iyonaga:2018vnu} by extending the 2-DOF construction from the scope of $k$-essence to the more general GLPV theory \cite{Gleyzes:2014dya}. In the unitary gauge, the action of the GLPV theory can be written as 
\begin{align}
	S_{\text{GLPV}} & =\int\text{d}t\text{d}^{3}xN\sqrt{h}\bigg[A_{2}+A_{3}K+A_{4}\left(\frac{2}{3}K^{2}-\hat{K}{}^{ij}\hat{K}{}_{ij}\right)+B_{4}R\nonumber \\
	& \ \ \ \ +A_{5}\left(\frac{2}{9}K^{3}-\hat{K}{}_{ij}\hat{K}{}^{ij}K+2\hat{K}{}_{ij}\hat{K}{}^{jk}\hat{K}{}_{k}^{i}\right)+B_{5}\bigg(R^{ij}\hat{K}{}_{ij}-\frac{1}{6}RK\bigg)\bigg],\label{GLPV}
\end{align}
where $A_{2},\cdots, A_{5}$, $B_{4}$ and $B_{5}$ are general functions of the time and lapse function. The extended cuscuton theory is generated from the above action \eqref{GLPV} by imposing special conditions. There are two branches of solutions for the extended cuscuton theory. For $A_{5}=0$, the first branch is
\begin{align}
	A_{5}  =0,\qquad  A_{4}  =-\frac{v_{4}N}{N+u_{4}}, \qquad A_{3}  =\frac{v_{3}}{N+u_{4}},
\end{align}
and
\begin{align}
	A_{2}  =u_{2}+\frac{v_{2}}{N}-\frac{3v_{3}^{2}}{8v_{4}N(N+u_{4})}, \qquad B_{4}=b_{0}+\frac{b_{1}}{N}, \qquad B_{5} &=0.
\end{align}
For $A_{5}\neq0$, the solutions are
\begin{align}
	A_{5}  =\frac{\pm N^{2}}{(\mu_{5}N+v_{5})^{2}}, \qquad A_{4}  =\frac{N(\mu_{4}N+v_{4})}{(\mu_{5}N+v_{5})^{2}}, \qquad A_{3}  =\mu_{3}\pm\frac{2(\mu_{4}N+\nu_{4})^{2}}{3(\mu_{5}N+\nu_{5})^{2}}, 
\end{align}
and
\begin{align}
	A_{2}  =\mu_{2}+\frac{v_{2}}{N}+\frac{2(\mu_{4}N+\nu_{4})^{3}}{9N(\mu_{5}N+\nu_{5})^{2}}, \qquad B_{4}  =b_{0}+\frac{b_{1}}{N}, \qquad B_{5}  =0.
\end{align}

Note that the linear extrinsic-curvature term, i.e., $\mathcal{L}_{1}$, is omitted in our theory in order to maintain agreement with general relativity. To compare our results with the extended cuscuton theory, we therefore focus on the subclass with $A_{3}=0$.

For the first branch, this requires $v_{3}=0$. For the second branch, we must require that the ratio $(\mu_{4}N+\nu_{4})/(\mu_{5}N+\nu_{5})$ be independent of the lapse function, in which case $A_{3}$ can be completely removed by taking $\mu_{3}$ as $\mu_{3}=\mp\frac{2}{3}\rho^{2}$. Under this assumption, the coefficients in the first branch reduce to
\begin{align}
	A_{5}  =0, \quad A_{4}  =-\frac{v_{4}N}{N+u_{4}}, \quad A_{3}  =0, \quad A_{2}  =\frac{v_{2}}{N}+u_{2}, \quad B_{4}  =\frac{b_{1}}{N}+b_{0}, \quad B_{5}  =0,
\end{align}
which we refer to as ``sub-class I''. The coefficients in the second branch reduce to
\begin{align}
	A_{5}  =\frac{\pm N^{2}}{(\mu_{5}N+v_{5})^{2}}, \quad A_{4}  =\frac{N\rho}{\mu_{5}N+v_{5}}, \quad A_{3}  =0, \quad A_{2}  =\frac{1}{N}\Big(v_{2}+\frac{2\nu_{4}\rho^{2}}{9}\Big)+\Big(\mu_{2}+\frac{2\mu_{4}\rho^{2}}{9}\Big),
\end{align}
and
\begin{align}
	B_{4} & =\frac{b_{1}}{N}+b_{0}, \quad B_{5}  =0,
\end{align}
which we refer to as ``sub-class II''.

It can be checked that sub-class I of the extended cuscuton theory can be recovered from the action $S^{\mathrm{A1}}$ \eqref{eq:S_A1} by choosing 
\begin{align}
	D_{3}=C_{6}  =0, \quad C_{3}  =b_{1}, \quad C_{4}  =b_{0}, \quad B_{1}  =v_{2}, \quad B_{2}  =u_{2},
\end{align}
and
\begin{align}
	C_{5}  =-2, \quad c_{1}^{(0;3,0)}  =0, \quad C_{1}  =\frac{2v_{4}}{u_{4}}, \quad C_{2}  =\frac{1}{u_{4}}.
\end{align}
That is,
\begin{align}
	S^{\text{sc-I}} & =\int\text{d}t\text{d}^{3}x\!N\sqrt{h}\Big[\frac{C_{1}N}{2(1+C_{2}N)}\Big(\frac{2}{3}K^{2}-\hat{K}^{ij}\hat{K}_{ij}\Big)+\Big(\frac{C_{3}}{N}+C_{4}\Big){}^{3}\!R+\Big(\frac{B_{1}}{N}+B_{2}\Big)\Big]\nonumber\\
	& =\int\text{d}t\text{d}^{3}x\!N\sqrt{h}\Big[\frac{v_{4}N}{(N+u_{4})}\Big(\frac{2}{3}K^{2}-\hat{K}^{ij}\hat{K}_{ij}\Big)+\Big(\frac{b_{1}}{N}+b_{0}\Big){}^{3}\!R+\Big(\frac{v_{2}}{N}+u_{2}\Big)\Big].
\end{align}
Similarly, sub-class II of the extended cuscuton theory can be recovered from the action $S^{\mathrm{A2}}$ \eqref{eq:S_A2} by choosing 
\begin{align}
	D_{3}  =0, \quad C_{3}  =b_{1}, \quad C_{4}  =b_{0}, \quad B_{1}  =v_{2}+\frac{2\nu_{4}\rho^{2}}{9}, \quad B_{2}  =\mu_{2}+\frac{2\mu_{4}\rho^{2}}{9},
\end{align}
and
\begin{align}
	C_{5}  =-2, \quad c_{1}^{(0;3,0)}  =\frac{D_{1}N^{2}}{(1+C_{2}N)^{2}}, \quad C_{1}  =\frac{2\rho}{\nu_{5}}, \quad C_{2}  =\frac{\mu_{5}}{v_{5}}, \quad D_{1}  =\pm\frac{1}{2v_{5}^{2}}.
\end{align}
That is,
\begin{align}
	S^{\text{sc-II}} & =\int\text{d}t\text{d}^{3}x\!N\sqrt{h}\Big[\frac{C_{1}N}{2(1+C_{2}N)}\Big(\frac{2}{3}K^{2}-\hat{K}{}^{ij}\hat{K}{}_{ij}\Big)+\Big(\frac{C_{3}}{N}+C_{4}\Big){}^{3}\!R+\Big(\frac{B_{1}}{N}+B_{2}\Big)\nonumber \\
	& \ \ \ \ +\frac{D_{1}N^{2}}{2(1+C_{2}N)^{2}}\Big(\frac{2}{9}K^{3}-\hat{K}{}_{ij}\hat{K}{}^{ij}K+2\hat{K}{}_{ij}\hat{K}{}^{jk}\hat{K}{}_{k}^{i}\Big)\Big]\nonumber \\
	& =\int\text{d}t\text{d}^{3}x\!N\sqrt{h}\Big[\frac{N\rho}{\mu_{5}N+v_{5}}\Big(\frac{2}{3}K^{2}-\hat{K}{}^{ij}\hat{K}{}_{ij}\Big)+\Big(\frac{b_{1}}{N}+b_{0}\Big){}^{3}\!R+\frac{1}{N}\Big(v_{2}+\frac{2\nu_{4}\rho^{2}}{9}\Big)+\Big(\mu_{2}+\frac{2\mu_{4}\rho^{2}}{9}\Big)\nonumber \\
	& \ \ \ \ +\frac{\pm N^{2}}{(\mu_{5}N+v_{5})^{2}}\Big(\frac{2}{9}K^{3}-\hat{K}{}_{ij}\hat{K}{}^{ij}K+2\hat{K}{}_{ij}\hat{K}{}^{jk}\hat{K}{}_{k}^{i}\Big)\Big].
\end{align}
We therefore conclude that our results are compatible with the extended cuscuton theory in the absence of the linear extrinsic-curvature term.

\subsection{Quadratic 2-DOF Theories}
\label{subsec:4.2}

In \cite{Gao:2019twq,Hu:2021yaq}, a 2-DOF theory quadratic in the extrinsic curvature was constructed, and it is free of the scalar DOF at the non-perturbative level. It can be checked that the Lagrangian of this quadratic 2-DOF theory is a special case of the action $S^{\mathrm{A1}}$ \eqref{eq:S_A1} with all coefficients in $d=0$ and $d=3$ set to zero, i.e.,
\begin{align}
	S^{(2)} & =\int\text{d}t\text{d}^{3}xN\sqrt{h}\bigg[\frac{C_{1}N}{C_{5}-2C_{2}N}K^{ij}K_{ij}+\frac{1}{3}\bigg(\frac{C_{1}N}{1+C_{2}N}-\frac{C_{1}N}{C_{5}-2C_{2}N}\bigg)K^{2}+\bigg(\frac{C_{3}}{N}+C_{4}\bigg)R\bigg],\label{eq:S(2)1}
\end{align}
which coincides exactly with the result in \cite{Hu:2021yaq}. One may further rewrite the action as 
\begin{align}
	S^{(2)} & =\int\text{d}t\text{d}^{3}xN\sqrt{h}\bigg[\frac{N}{\frac{C_{5}}{C_{1}}+(-2\frac{C_{2}}{C_{1}})N}K^{ij}K_{ij}-\frac{1}{3}\bigg(\frac{2N}{-\frac{2}{C_{1}}+(-2\frac{C_{2}}{C_{1}})N}+\frac{N}{\frac{C_{5}}{C_{1}}+(-2\frac{C_{2}}{C_{1}})N}\bigg)K^{2}\nonumber \\
	& \ \ \ \ +\bigg(\frac{C_{3}}{N}+C_{4}\bigg)R\bigg].\label{eq:action of d=2}
\end{align}
Such a quadratic 2-DOF theory was first constructed on the basis of the Hamiltonian constraint analysis in \cite{Gao:2019twq}, where the action is written as
\begin{align}
	S^{(\text{quad})} & =\int\text{d}t\text{d}^{3}xN\sqrt{h}\bigg[\frac{N}{\beta_{2}+\beta_{5}N}K^{ij}K_{ij}-\frac{1}{3}\bigg(\frac{2N}{\beta_{1}+\beta_{5}N}+\frac{N}{\beta_{2}+\beta_{5}N}\bigg)K^{2}+\bigg(\frac{\rho_{3}}{N}+\rho_{4}+\rho_{2}\bigg)R+\rho_{1}\bigg],
\end{align}
which is nothing but the action \eqref{eq:action of d=2} after redefining the coefficients as\footnote{Note the integration constant $\beta_{5}$ is set to unity in \cite{Gao:2019twq}.}
\begin{align}
	\beta_{1}  =-\frac{2}{C_{1}}, \quad \beta_{2}  =\frac{C_{5}}{C_{1}}, \quad \rho_{3} =C_{3}, \quad \rho_{4}+\rho_{2}  =C_{4}, \quad \beta_{5} & =-2\frac{C_{2}}{C_{1}}.
\end{align}
We therefore conclude that we have reproduced the quadratic 2-DOF theory, which propagates no scalar mode at the fully nonlinear level, by using a perturbative approach up to cubic order in perturbations. This matching also illustrates the ability of the perturbative method to construct 2-DOF theories from a finite-order perturbation analysis.

\section{Conclusion}
\label{sec:5}

A defining feature of spatially covariant gravity (SCG) is the breaking of spacetime diffeomorphism invariance. This allows SCG to evade Lovelock's theorem and thereby provides a genuine framework for modifying general relativity. In \cite{Gao:2014fra, Gao:2014soa}, a class of SCG theories propagating one scalar and two tensor degrees of freedom (DOFs) was systematically constructed. In the present work, we have investigated the conditions under which the scalar mode becomes nondynamical, so that only 2 DOFs propagate. Although Hamiltonian formulations of 2-DOF SCG theories have been developed in \cite{Gao:2019twq, Yao:2020tur, Yao:2023qjd}, the corresponding conditions are often difficult to translate into explicit Lagrangian form. For this reason, we adopted a fully Lagrangian construction based on perturbation analysis.

Using the perturbative method developed in \cite{Gao:2019lpz,Hu:2021yaq}, we derived the necessary 2-DOF conditions for the polynomial-type SCG Lagrangian \eqref{eq:action of d=0, d=2 and d=3}, which consists of monomials with $d=0$ as in \eqref{eq:L0}, $d=2$ as in \eqref{eq:L2}, and $d=3$ as in \eqref{eq:L3}. We found that the conditions obtained from linear perturbations are not sufficient. At the same time, explicit examples of SCG, both with a propagating lapse function (see \cite{Gao:2019lpz} for details) and without one (see Sec. \ref{subsec:4.2} for an illustration), indicate that conditions derived up to cubic order in the perturbative Lagrangian can provide additional constraints capable of further removing the scalar degree of freedom.

The main goal of this work has been to identify the 2-DOF conditions that arise at cubic order in perturbations for polynomial-type SCG theories. Building on \cite{Hu:2021yaq}, we extended the perturbative expansion around a Friedmann--Robertson--Walker (FRW) background to the next order. In Sec. \ref{sec:2}, we briefly reviewed the second-order conditions for eliminating the extra mode. The background evolution equations \eqref{eq:bgeq1} and \eqref{eq:bgeq2} were used to simplify the subsequent perturbative expressions. After substituting the solutions for the auxiliary fields $A$ and $B$, the degeneracy condition for $\zeta$ at quadratic order follows from the effective Lagrangian and is explicitly given in \eqref{eq:solution for b2} through \eqref{eq:solution for omega3}. In particular, the allowed coefficient functions at this stage split into two main branches, referred to as Case I and Case II.

The main analysis was carried out in Sec. \ref{sec:3}. After tuning the coefficients so that the quadratic-order conditions are satisfied, we examined the cubic terms governing the temporal evolution and organized them according to the number of derivatives. In Case I, this led to five explicit actions: $S^{\mathrm{A1}}$ in \eqref{eq:S_A1}, $S^{\mathrm{A2}}$ in \eqref{eq:S_A2}, $S^{\mathrm{B1}}$ in \eqref{eq:S_B1}, $S^{\mathrm{B2}}$ in \eqref{eq:S_B2}, and $S^{\mathrm{C}}$ in \eqref{eq:S_C}. These actions arise by imposing the cubic-order degeneracy conditions on the quadratic-order solutions, and thus constitute the full set of candidate theories that survive within the polynomial class considered here up to cubic order in perturbations. By contrast, in Case II we analyzed the corresponding conditions in detail (see Appendix \ref{sec:Degeneracy Conditions: case II}) and found that no admissible solution survives. Thus, once the cubic-order requirements are imposed, the quadratic-order branch Case II is completely ruled out. Moreover, among the five actions $S^{\mathrm{A1}},\cdots, S^{\mathrm{C}}$, only $S^{\mathrm{A1}}$ and $S^{\mathrm{A2}}$ are compatible with general relativity in the low-energy limit. This further singles out these two theories as the most physically viable candidates in our construction, whereas the remaining three actions still satisfy the perturbative degeneracy conditions but do not reproduce the desired infrared behavior.

In Sec. \ref{sec:4}, we further examined the relation between our construction and the extended cuscuton theory, as well as the quadratic 2-DOF theories, and showed that both are compatible with our framework. More specifically, these comparisons clarify how previously known 2-DOF theories can be embedded into, or consistently interpreted within, the broader class of polynomial-type SCG theories studied here. In this sense, our construction not only provides new candidate actions, but also offers a unified perturbative perspective on the relation among different realizations of 2-DOF spatially covariant gravity.

Overall, our analysis has produced five candidate 2-DOF actions that are free of scalar propagation at least up to cubic order in perturbation theory. The results therefore provide a concrete and systematic step toward identifying scalar-less sectors of polynomial-type SCG directly in the Lagrangian language. Nevertheless, it remains open whether the necessary conditions derived here are also sufficient to eliminate the scalar mode completely beyond the perturbative order considered. In principle, one may extend the same analysis to quartic and higher orders in perturbations, or use independent methods to verify the complete removal of the scalar mode at the nonperturbative level. It would also be worthwhile to investigate in more detail the physical implications of the viable branches, especially the low-energy behavior of $S^{\mathrm{A1}}$ and $S^{\mathrm{A2}}$ and their possible phenomenological significance. We leave these questions for future work.

\acknowledgments
This work was partly supported by National Natural Science Foundation of China (NSFC) under Grants No.12475068, No.11975020 and No.12547120, and the Guangdong Basic and Applied Basic Research Foundation under Grant No. 2025A1515012977.

\appendix

\section{Explicit expressions of the coefficients in Case I}
\label{sec:Explicit form of the coefficients in case I}

The explicit expressions of $\Delta_{\dot{\zeta}\partial^{2}\zeta\partial^{2}\zeta,i}^{\text{I}}$ with $i=0,...,1$ in (\ref{eq:DeltaI}) are 
\begin{align}
\Delta_{\dot{\zeta}\partial^{2}\zeta\partial^{2}\zeta,4}^{\text{I}} & =-9C_{2}H^{4}\bar{N}^{3}\bigg[-8D_{4}^{2}\omega_{3}^{2}+C_{2}^{2}D_{1}\tilde{c}_{1}^{(1;1,0)}\bar{N}^{2}\bigl(2\tilde{c}_{1}^{(1;1,0)\prime}\omega_{3}+\tilde{c}_{1}^{(1;1,0)}(\omega_{3}-\omega_{3}^{\prime})\bigr)\nonumber \\
 & \ \ \ \ +C_{2}D_{1}\tilde{c}_{1}^{(1;1,0)}\bar{N}(2\tilde{c}_{1}^{(1;1,0)}\omega_{3}+2\tilde{c}_{1}^{(1;1,0)\prime}\omega_{3}-\tilde{c}_{1}^{(1;1,0)}\omega_{3}^{\prime})\bigg],
\end{align}
\begin{align}
\Delta_{\dot{\zeta}\partial^{2}\zeta\partial^{2}\zeta,3}^{\text{I}} & =-3C_{2}H^{3}\bar{N}^{3}\bigg\{ C_{1}C_{2}^{3}\tilde{c}_{1}^{(1;1,0)}\bar{N}^{2}\bigg(2\tilde{c}_{1}^{(1;1,0)\prime}\omega_{3}+\tilde{c}_{1}^{(1;1,0)}(\omega_{3}-\omega_{3}^{\prime})\bigg)\nonumber \\
 & \ \ \ \ +4\bigg(-4D_{4}^{2}\omega_{2}\omega_{3}+3C_{4}D_{1}(2\tilde{c}_{1}^{(1;1,0)}\omega_{3}+\tilde{c}_{1}^{(1;1,0)\prime}\omega_{3}-\tilde{c}_{1}^{(1;1,0)}\omega_{3}^{\prime})\bigg)\nonumber \\
 & \ \ \ \ +C_{2}\bigg[C_{1}\tilde{c}_{1}^{(1;1,0)}(2\tilde{c}_{1}^{(1;1,0)}\omega_{3}+2\tilde{c}_{1}^{(1;1,0)\prime}\omega_{3}-\tilde{c}_{1}^{(1;1,0)}\omega_{3}^{\prime})\nonumber \\
 & \ \ \ \ +D_{1}\bar{N}\bigg(\tilde{c}_{1}^{(1;1,0)2}(\omega_{2}-\omega_{2}^{\prime})+24C_{4}\tilde{c}_{1}^{(1;1,0)\prime}\omega_{3}+2\tilde{c}_{1}^{(1;1,0)}(\tilde{c}_{1}^{(1;1,0)\prime}\omega_{2}+18C_{4}\omega_{3}-12C_{4}\omega_{3}^{\prime})\bigg)\bigg]\nonumber \\
 & \ \ \ \ +C_{2}^{2}\bar{N}\bigg[C_{1}\tilde{c}_{1}^{(1;1,0)}(3\tilde{c}_{1}^{(1;1,0)}\omega_{3}+4\tilde{c}_{1}^{(1;1,0)\prime}\omega_{3}-2\tilde{c}_{1}^{(1;1,0)}\omega_{3}^{\prime})\nonumber \\
 & \ \ \ \ +D_{1}\bar{N}\bigg(\tilde{c}_{1}^{(1;1,0)2}(\omega_{2}-\omega_{2}^{\prime})+12C_{4}\tilde{c}_{1}^{(1;1,0)\prime}\omega_{3}+2\tilde{c}_{1}^{(1;1,0)}(\tilde{c}_{1}^{(1;1,0)\prime}\omega_{2}+6C_{4}\omega_{3}-6C_{4}\omega_{3}^{\prime})\bigg)\bigg]\bigg\},
\end{align}
\begin{align}
\Delta_{\dot{\zeta}\partial^{2}\zeta\partial^{2}\zeta,2}^{\text{I}} & =-H^{2}\bar{N}\bigg\{72C_{2}^{5}C_{4}^{2}\bar{N}^{4}\omega_{3}^{2}+36C_{4}^{2}D_{1}\bar{N}(2\omega_{3}-\omega_{3}^{\prime})\nonumber \\
 & +C_{2}^{2}\bar{N}\bigg[C_{1}\bar{N}\bigg(\tilde{c}_{1}^{(1;1,0)2}(\omega_{2}-\omega_{2}^{\prime})+36C_{4}\tilde{c}_{1}^{(1;1,0)\prime}\omega_{3}+2\tilde{c}_{1}^{(1;1,0)}(\tilde{c}_{1}^{(1;1,0)\prime}\omega_{2}+30C_{4}\omega_{3}-18C_{4}\omega_{3}^{\prime})\bigg)\nonumber \\
 & +12C_{4}\bigg(24C_{4}\omega_{3}^{2}+D_{1}\bar{N}^{2}(2\tilde{c}_{1}^{(1;1,0)}\omega_{2}+2\tilde{c}_{1}^{(1;1,0)\prime}\omega_{2}-2\tilde{c}_{1}^{(1;1,0)}\omega_{2}^{\prime}+12C_{4}\omega_{3}-9C_{4}\omega_{3}^{\prime})\bigg)\bigg]\nonumber \\
 & +C_{2}^{4}\bar{N}^{3}\bigg[288C_{4}^{2}\omega_{3}^{2}+C_{1}\bar{N}\bigg(\tilde{c}_{1}^{(1;1,0)2}(\omega_{2}-\omega_{2}^{\prime})\nonumber \\
 & +12C_{4}\tilde{c}_{1}^{(1;1,0)\prime}\omega_{3}+2\tilde{c}_{1}^{(1;1,0)}(\tilde{c}_{1}^{(1;1,0)\prime}\omega_{2}+6C_{4}\omega_{3}-6C_{4}\omega_{3}^{\prime})\bigg)\bigg]\nonumber \\
 & +2C_{2}^{3}\bar{N}^{2}\bigg[C_{1}\bar{N}\bigg(\tilde{c}_{1}^{(1;1,0)2}(\omega_{2}-\omega_{2}^{\prime})+18C_{4}\tilde{c}_{1}^{(1;1,0)\prime}\omega_{3}+2\tilde{c}_{1}^{(1;1,0)}(\tilde{c}_{1}^{(1;1,0)\prime}\omega_{2}+12C_{4}\omega_{3}-9C_{4}\omega_{3}^{\prime})\bigg)\nonumber \\
 & +6C_{4}\bigg(36C_{4}\omega_{3}^{2}+D_{1}\bar{N}^{2}(\tilde{c}_{1}^{(1;1,0)}\omega_{2}+\tilde{c}_{1}^{(1;1,0)\prime}\omega_{2}-\tilde{c}_{1}^{(1;1,0)}\omega_{2}^{\prime}+3C_{4}\omega_{3}-3C_{4}\omega_{3}^{\prime})\bigg)\bigg]\nonumber \\
 & +4C_{2}\bigg[-2D_{4}^{2}\bar{N}^{2}\omega_{2}^{2}+9C_{4}^{2}\bigg(2\omega_{3}^{2}+D_{1}\bar{N}^{2}(5\omega_{3}-3\omega_{3}^{\prime})\bigg)\nonumber \\
 & +3C_{4}\bar{N}\bigg(D_{1}\bar{N}(\tilde{c}_{1}^{(1;1,0)}\omega_{2}+\tilde{c}_{1}^{(1;1,0)\prime}\omega_{2}-\tilde{c}_{1}^{(1;1,0)}\omega_{2}^{\prime})+C_{1}(2\tilde{c}_{1}^{(1;1,0)}\omega_{3}+\tilde{c}_{1}^{(1;1,0)\prime}\omega_{3}-\tilde{c}_{1}^{(1;1,0)}\omega_{3}^{\prime})\bigg)\bigg]\bigg\},
\end{align}
\begin{align}
\Delta_{\dot{\zeta}\partial^{2}\zeta\partial^{2}\zeta,1}^{\text{I}} & =-4C_{4}H\bar{N}(1+C_{2}\bar{N})^{3}\bigg[C_{1}C_{2}\bar{N}\bigg(\tilde{c}_{1}^{(1;1,0)\prime}\omega_{2}+\tilde{c}_{1}^{(1;1,0)}(\omega_{2}-\omega_{2}^{\prime})\bigg)\nonumber \\
 & \ \ \ \ +3C_{4}\bigg(D_{1}\bar{N}(\omega_{2}-\omega_{2}^{\prime})+C_{1}(2+C_{2}\bar{N})\omega_{3}\nonumber \\
 & \ \ \ \ +4C_{2}(1+C_{2}\bar{N})\omega_{2}\omega_{3}-C_{1}(1+C_{2}\bar{N})\omega_{3}^{\prime}\bigg)\bigg],
\end{align}
and
\begin{align}
\Delta_{\dot{\zeta}\partial^{2}\zeta\partial^{2}\zeta,0}^{\text{I}} & =-4C_{4}^{2}\bar{N}(1+C_{2}\bar{N})^{4}\bigg(2C_{2}\omega_{2}^{2}+C_{1}(\omega_{2}-\omega_{2}^{\prime})\bigg)
\end{align}
with $\tilde{c}_{1}^{(1;1,0)}\coloneqq c_{1}^{(1;1,0)}+2D_{3}$.

\section{Degeneracy analysis in Case II}
\label{sec:Degeneracy Conditions: case II}

In this appendix, we present the procedure used to analyze the degeneracy conditions in Case II. 
From (\ref{eq:Delta1}), if $\tilde{f}_{3} \neq 0$, then $\Delta_{1}$ contains $\partial^2$, and the corresponding solutions for $A$ and $B$ therefore involve nonlocal operators. Accordingly, we divide the analysis into two subcases, depending on whether $\tilde{f}_{3}$ vanishes.

\subsection{$\tilde{f}_{3} \neq 0$}

We first consider the subcase with $\tilde{f}_{3}\neq 0$.

Since the calculations in this part rely heavily on the operator $\Delta_{1}$ in \eqref{eq:Delta1}, it is useful to recast the relevant expressions into a form that makes the dependence on $\Delta_{1}^{-1}$ explicit. In particular, the solutions for $A$ and $B$ can be rewritten equivalently as
\begin{align}
	A & =\frac{1}{\Delta_{1}}2(f_{3}-3\tilde{f}_{3})\bar{N}(1+C_{2}\bar{N})(C_{1}+C_{1}C_{2}\bar{N}+3D_{1}H\bar{N})a^{2}\dot{\zeta}\nonumber \\
	& \ \ \ \ -\frac{1}{\Delta_{1}}\frac{2\tilde{f}_{3}(1+C_{2}\bar{N})^{3}}{C_{2}(f_{3}-3\tilde{f}_{3})H\bar{N}}\bigg[C_{2}f_{3}H\bar{N}(c_{1}^{(1;1,0)}+2J_{3})\nonumber \\
	& \ \ \ \ \ \ \ \ +4f_{3}(h_{2}+h_{2}^{\prime}+HJ_{3}^{\prime})(1+C_{2}\bar{N})-6\tilde{f}_{3}(h_{2}+h_{2}^{\prime}+HJ_{3}^{\prime})(1+C_{2}\bar{N})\bigg]\partial^{2}\zeta,
\end{align}
and
\begin{align}
	B & =\frac{1}{\Delta_{1}}f_{3}^{2}(1+C_{2}\bar{N})^{3}a\dot{\zeta}\nonumber \\
	& \ \ \ \ +\frac{1}{\Delta_{1}}\frac{2f_{3}(1+C_{2}\bar{N})^{2}a}{3(f_{3}-3\tilde{f}_{3})\tilde{f}_{3}}\bigg[-2C_{2}f_{3}^{2}H(c_{1}^{(1;1,0)}+2J_{3})\bar{N}\nonumber \\
	& \ \ \ \ +9C_{2}f_{3}\tilde{f}_{3}H(c_{1}^{(1;1,0)}+2J_{3})\bar{N}+18\tilde{f}_{3}^{2}(h_{2}+h_{2}^{\prime}+HJ_{3}^{\prime})(1+C_{2}\bar{N})\bigg]\zeta,
\end{align}
which are both ``proportional'' to $\Delta_{1}^{-1}$. 
Before substituting these solutions into the cubic Lagrangian for the perturbations $\mathcal{L}_{3}(\zeta,A,B)$, we also rewrite $\zeta$ in the analogous form
\begin{align}
	\zeta & =\frac{1}{\Delta_{1}}f_{3}\tilde{f}_{3}(1+C_{2}\bar{N})^{3}\partial^{2}\zeta+\frac{1}{\Delta_{1}}2C_{2}(f_{3}-3\tilde{f}_{3})H\bar{N}^{2}(C_{1}+C_{1}C_{2}\bar{N}+3D_{1}H\bar{N})a^{2}\zeta,
\end{align}
so that $\zeta$, $A$, and $B$ are all formally acted on by the inverse operator $\Delta_{1}^{-1}$. After substituting the solutions for $A$ and $B$, all terms in the cubic Lagrangian become ``proportional'' to $\Delta_{1}^{-3}$, and the assumption $\tilde{f}_{3}\neq 0$ guarantees that $\Delta_{1}$ never vanishes. Since all terms in the cubic Lagrangian share the same ``denominator'' $\Delta_{1}^{3}$, we may focus only on the remaining parts of the coefficients in the degeneracy analysis.

To keep the discussion parallel to that in Case I, we classify the conditions by the total number of temporal derivatives appearing in each term.

\subsubsection{Terms with three temporal derivatives}

The terms in the cubic action involving three temporal derivatives take the form
\begin{align}
	\mathcal{L}_{3,t3}^{\text{II}} & \simeq\hat{\mathcal{C}}_{\partial^{j}\partial^{i}\dot{\zeta}\partial_{k}\partial_{i}\dot{\zeta}\partial^{k}\partial_{j}\dot{\zeta}}^{\text{II}}\partial^{j}\partial^{i}\dot{\zeta}\partial_{k}\partial_{i}\dot{\zeta}\partial^{k}\partial_{j}\dot{\zeta}+\hat{\mathcal{C}}_{\partial^{2}\dot{\zeta}\partial^{2}\dot{\zeta}\partial^{2}\dot{\zeta}}^{\text{II}}\partial^{2}\dot{\zeta}\partial^{2}\dot{\zeta}\partial^{2}\dot{\zeta}\nonumber \\
	& \ \ \ \ +\hat{\mathcal{C}}_{\partial^{2}\dot{\zeta}\partial^{i}\partial^{j}\dot{\zeta}\partial_{i}\partial_{j}\dot{\zeta}}^{\text{II}}\partial^{2}\dot{\zeta}\partial^{i}\partial^{j}\dot{\zeta}\partial_{i}\partial_{j}\dot{\zeta}\nonumber \\
	& \ \ \ \ +\hat{\mathcal{C}}_{\partial^{i}\dot{\zeta}\partial_{j}\partial_{i}\dot{\zeta}\partial^{j}\dot{\zeta}}^{\text{II}}\partial^{i}\dot{\zeta}\partial^{j}\dot{\zeta}\partial_{j}\partial_{i}\dot{\zeta}+\hat{\mathcal{C}}_{\partial^{2}\dot{\zeta}\partial^{i}\dot{\zeta}\partial_{i}\dot{\zeta}}^{\text{II}}\partial^{2}\dot{\zeta}\partial^{i}\dot{\zeta}\partial_{i}\dot{\zeta}\nonumber \\
	& \ \ \ \ +\hat{\mathcal{C}}_{\partial^{2}\dot{\zeta}\partial^{2}\dot{\zeta}\dot{\zeta}}^{\text{II}}\partial^{2}\dot{\zeta}\partial^{2}\dot{\zeta}\dot{\zeta}+\hat{\mathcal{C}}_{\partial^{i}\partial^{j}\dot{\zeta}\partial_{i}\partial_{j}\dot{\zeta}\dot{\zeta}}^{\text{II}}\partial^{i}\partial^{j}\dot{\zeta}\partial_{i}\partial_{j}\dot{\zeta}\dot{\zeta}\nonumber \\
	& \ \ \ \ +\hat{\mathcal{C}}_{\partial^{2}\dot{\zeta}\dot{\zeta}\dot{\zeta}}^{\text{II}}\partial^{2}\dot{\zeta}\dot{\zeta}\dot{\zeta}+\hat{\mathcal{C}}_{\partial^{i}\dot{\zeta}\partial_{i}\dot{\zeta}\dot{\zeta}}^{\text{II}}\partial^{i}\dot{\zeta}\partial_{i}\dot{\zeta}\dot{\zeta}. \label{calL_II3t3}
\end{align}
As in Case I, only terms with the same number of spatial derivatives can cancel one another after integrations by parts. We therefore analyze the different spatial-derivative sectors separately.

For terms with six spatial derivatives, one of the three terms in the first two lines of (\ref{calL_II3t3}) can be reduced by integration by parts, leaving two conditions:
\begin{align}
	-\frac{1}{2}\hat{\mathcal{C}}_{\partial^{j}\partial^{i}\dot{\zeta}\partial_{k}\partial_{i}\dot{\zeta}\partial^{k}\partial_{j}\dot{\zeta}}^{\text{II}}+\hat{\mathcal{C}}_{\partial^{2}\dot{\zeta}\partial^{2}\dot{\zeta}\partial^{2}\dot{\zeta}}^{\text{II}} & =0,\\
	\frac{3}{2}\hat{\mathcal{C}}_{\partial^{j}\partial^{i}\dot{\zeta}\partial_{k}\partial_{i}\dot{\zeta}\partial^{k}\partial_{j}\dot{\zeta}}^{\text{II}}+\hat{\mathcal{C}}_{\partial^{2}\dot{\zeta}\partial^{i}\partial^{j}\dot{\zeta}\partial_{i}\partial_{j}\dot{\zeta}}^{\text{II}} & =0.
\end{align}
From these, we solve for the coefficient $c_{3}^{(0;3,0)}$ as
\begin{align}
	c_{3}^{(0;3,0)} & =-\frac{3D_{1}(f_{3}-3\tilde{f}_{3})^{2}(f_{3}-2\tilde{f}_{3})\bar{N}^{2}}{2f_{3}^{3}(1+C_{2}\bar{N})^{2}}. \label{sol_IIt3s6}
\end{align}

For terms with four spatial derivatives, there are two integration-by-parts relations among four terms, leaving two independent conditions:
\begin{align}
	-\frac{1}{2}\hat{\mathcal{C}}_{\partial^{i}\dot{\zeta}\partial_{j}\partial_{i}\dot{\zeta}\partial^{j}\dot{\zeta}}^{\text{II}}+\hat{\mathcal{C}}_{\partial^{2}\dot{\zeta}\partial^{i}\dot{\zeta}\partial_{i}\dot{\zeta}}^{\text{II}}+\frac{3}{2}\hat{\mathcal{C}}_{\partial^{i}\partial^{j}\dot{\zeta}\partial_{i}\partial_{j}\dot{\zeta}\dot{\zeta}}^{\text{II}} & =0,\label{cond_IIt3s4_1} \\
	\hat{\mathcal{C}}_{\partial^{i}\partial^{j}\dot{\zeta}\partial_{i}\partial_{j}\dot{\zeta}\dot{\zeta}}^{\text{II}}+\hat{\mathcal{C}}_{\partial^{2}\dot{\zeta}\partial^{2}\dot{\zeta}\dot{\zeta}}^{\text{II}} & =0. \label{cond_IIt3s4_2}
\end{align}
The first condition (\ref{cond_IIt3s4_1}) gives
\begin{align}
	c_{4}^{(0;3,0)} & =\frac{f_{3}^{\prime}(f_{3}-\tilde{f}_{3})}{2f_{3}}+\frac{1}{20}\frac{C_{2}(7f_{3}-\tilde{f}_{3})\bar{N}}{1+C_{2}\bar{N}}. \label{sol_IIt3s4}
\end{align}
After imposing this result, the second condition (\ref{cond_IIt3s4_2}) is automatically satisfied.

For terms with two spatial derivatives, there is only one condition,
\begin{align}
	-\frac{1}{2}\hat{\mathcal{C}}_{\partial^{i}\dot{\zeta}\partial_{i}\dot{\zeta}\dot{\zeta}}^{\text{II}}+\hat{\mathcal{C}}_{\partial^{2}\dot{\zeta}\dot{\zeta}\dot{\zeta}}^{\text{II}}=0 ,
\end{align}
which implies
\begin{equation}
	f_{3}^{\prime}=0. \label{sol_IIt3s2}
\end{equation}

Combining (\ref{sol_IIt3s6}), (\ref{sol_IIt3s4}), and (\ref{sol_IIt3s2}), we summarize the solutions obtained from the sector with three temporal derivatives as
\begin{align}
	c_{3}^{(0;3,0)} & =-\frac{3D_{1}(f_{3}-3\tilde{f}_{3})^{2}(f_{3}-2\tilde{f}_{3})N^{2}}{2f_{3}^{3}(1+C_{2}N)^{2}},\\
	c_{4}^{(0;3,0)} & =\frac{1}{20}\frac{C_{2}(7f_{3}-\tilde{f}_{3})N}{1+C_{2}N},\\
	f_{3} & =D_{6}.
\end{align}

\subsubsection{Terms with two temporal derivatives}

We next consider the terms in the cubic Lagrangian involving two temporal derivatives:
\begin{align}
	\mathcal{L}_{3,t2}^{\text{II}} & \simeq\hat{\mathcal{C}}_{\partial^{2}\zeta\partial^{2}\dot{\zeta}\partial^{2}\dot{\zeta}}^{\text{II}}\partial^{2}\zeta\partial^{2}\dot{\zeta}\partial^{2}\dot{\zeta}+\hat{\mathcal{C}}_{\partial^{i}\partial^{j}\zeta\partial^{k}\partial_{i}\dot{\zeta}\partial_{k}\partial_{j}\dot{\zeta}}^{\text{II}}\partial^{i}\partial^{j}\zeta\partial^{k}\partial_{i}\dot{\zeta}\partial_{k}\partial_{j}\dot{\zeta}\nonumber \\
	& \ \ \ \ +\hat{\mathcal{C}}_{\partial^{i}\partial^{j}\zeta\partial_{i}\partial_{j}\dot{\zeta}\partial^{2}\dot{\zeta}}^{\text{II}}\partial^{i}\partial^{j}\zeta\partial_{i}\partial_{j}\dot{\zeta}\partial^{2}\dot{\zeta}+\hat{\mathcal{C}}_{\partial^{2}\zeta\partial^{j}\partial^{i}\dot{\zeta}\partial_{i}\partial_{j}\dot{\zeta}}^{\text{II}}\partial^{2}\zeta\partial^{j}\partial^{i}\dot{\zeta}\partial_{i}\partial_{j}\dot{\zeta}\nonumber \\
	& \ \ \ \ +\hat{\mathcal{C}}_{\partial^{j}\partial^{2}\zeta\partial_{j}\dot{\zeta}\partial^{2}\dot{\zeta}}^{\text{II}}\partial^{j}\partial^{2}\zeta\partial_{j}\dot{\zeta}\partial^{2}\dot{\zeta}+\hat{\mathcal{C}}_{\partial^{j}\partial^{2}\zeta\partial_{k}\partial_{j}\dot{\zeta}\partial^{k}\dot{\zeta}}^{\text{II}}\partial^{j}\partial^{2}\zeta\partial_{k}\partial_{j}\dot{\zeta}\partial^{k}\dot{\zeta}\nonumber \\
	& \ \ \ \ +\hat{\mathcal{C}}_{\partial^{4}\zeta\partial^{2}\dot{\zeta}\dot{\zeta}}^{\text{II}}\partial^{4}\zeta\partial^{2}\dot{\zeta}\dot{\zeta}+\hat{\mathcal{C}}_{\partial^{i}\partial^{j}\partial^{2}\zeta\partial_{i}\partial_{j}\dot{\zeta}\dot{\zeta}}^{\text{II}}\partial^{i}\partial^{j}\partial^{2}\zeta\partial_{i}\partial_{j}\dot{\zeta}\dot{\zeta}\nonumber \\
	& \ \ \ \ +\hat{\mathcal{C}}_{\zeta\partial^{j}\partial^{i}\dot{\zeta}\partial_{i}\partial_{j}\dot{\zeta}}^{\text{II}}\zeta\partial^{j}\partial^{i}\dot{\zeta}\partial_{i}\partial_{j}\dot{\zeta}+\hat{\mathcal{C}}_{\zeta\partial^{2}\dot{\zeta}\partial^{2}\dot{\zeta}}^{\text{II}}\zeta\partial^{2}\dot{\zeta}\partial^{2}\dot{\zeta}+\hat{\mathcal{C}}_{\partial^{2}\zeta\partial^{i}\dot{\zeta}\partial_{i}\dot{\zeta}}^{\text{II}}\partial^{2}\zeta\partial^{i}\dot{\zeta}\partial_{i}\dot{\zeta}\nonumber \\
	& \ \ \ \ +\hat{\mathcal{C}}_{\partial^{i}\zeta\partial^{j}\dot{\zeta}\partial_{i}\partial_{j}\dot{\zeta}}^{\text{II}}\partial^{i}\zeta\partial^{j}\dot{\zeta}\partial_{i}\partial_{j}\dot{\zeta}+\hat{\mathcal{C}}_{\partial^{j}\partial^{i}\zeta\partial_{i}\dot{\zeta}\partial_{j}\dot{\zeta}}^{\text{II}}\partial^{j}\partial^{i}\zeta\partial_{i}\dot{\zeta}\partial_{j}\dot{\zeta}+\hat{\mathcal{C}}_{\partial^{i}\zeta\partial^{2}\dot{\zeta}\partial_{i}\dot{\zeta}}^{\text{II}}\partial^{i}\zeta\partial^{2}\dot{\zeta}\partial_{i}\dot{\zeta}\nonumber \\
	& \ \ \ \ +\hat{\mathcal{C}}_{\partial^{2}\zeta\partial^{2}\dot{\zeta}\dot{\zeta}}^{\text{II}}\partial^{2}\zeta\partial^{2}\dot{\zeta}\dot{\zeta}+\hat{\mathcal{C}}_{\partial^{i}\partial^{j}\zeta\partial_{i}\partial_{j}\dot{\zeta}\dot{\zeta}}^{\text{II}}\partial^{i}\partial^{j}\zeta\partial_{i}\partial_{j}\dot{\zeta}\dot{\zeta}+\hat{\mathcal{C}}_{\partial^{i}\partial^{2}\zeta\partial_{i}\dot{\zeta}\dot{\zeta}}^{\text{II}}\partial^{i}\partial^{2}\zeta\partial_{i}\dot{\zeta}\dot{\zeta}+\hat{\mathcal{C}}_{\partial^{4}\zeta\dot{\zeta}\dot{\zeta}}^{\text{II}}\partial^{4}\zeta\dot{\zeta}^{2}\nonumber \\
	& \ \ \ \ +\hat{\mathcal{C}}_{\zeta\partial^{i}\dot{\zeta}\partial_{i}\dot{\zeta}}^{\text{II}}\zeta\partial^{i}\dot{\zeta}\partial_{i}\dot{\zeta}+\hat{\mathcal{C}}_{\zeta\partial^{2}\dot{\zeta}\dot{\zeta}}^{\text{II}}\zeta\partial^{2}\dot{\zeta}\dot{\zeta}+\hat{\mathcal{C}}_{\partial^{2}\zeta\dot{\zeta}\dot{\zeta}}^{\text{II}}\partial^{i}\zeta\partial_{i}\dot{\zeta}\dot{\zeta}+\hat{\mathcal{C}}_{\partial^{2}\zeta\dot{\zeta}\dot{\zeta}}^{\text{II}}\partial^{2}\zeta\dot{\zeta}^{2}.
\end{align}

We begin with the sector containing six spatial derivatives. 
Among the eight terms with six spatial derivatives, three can be reduced by integrations by parts, leaving the five conditions
\begin{align}
	\hat{\mathcal{C}}_{\partial^{2}\zeta\partial^{2}\dot{\zeta}\partial^{2}\dot{\zeta}}^{\text{II}}+\hat{\mathcal{C}}_{\partial^{2}\zeta\partial^{j}\partial^{i}\dot{\zeta}\partial_{i}\partial_{j}\dot{\zeta}}^{\text{II}} & =0, \label{cond_IIt2s6_1}\\
	\hat{\mathcal{C}}_{\partial^{i}\partial^{j}\zeta\partial_{i}\partial_{j}\dot{\zeta}\partial^{2}\dot{\zeta}}^{\text{II}}+\hat{\mathcal{C}}_{\partial^{2}\zeta\partial^{j}\partial^{i}\dot{\zeta}\partial_{i}\partial_{j}\dot{\zeta}}^{\text{II}} & =0, \label{cond_IIt2s6_2}\\
	\hat{\mathcal{C}}_{\partial^{j}\partial^{2}\zeta\partial_{j}\dot{\zeta}\partial^{2}\dot{\zeta}}^{\text{II}}+\hat{\mathcal{C}}_{\partial^{j}\partial^{2}\zeta\partial_{k}\partial_{j}\dot{\zeta}\partial^{k}\dot{\zeta}}^{\text{II}} & =0, \label{cond_IIt2s6_3}\\
	\hat{\mathcal{C}}_{\partial^{2}\zeta\partial^{2}\dot{\zeta}\partial^{2}\dot{\zeta}}^{\text{II}}-\frac{1}{2}\hat{\mathcal{C}}_{\partial^{i}\partial^{j}\zeta\partial^{k}\partial_{i}\dot{\zeta}\partial_{k}\partial_{j}\dot{\zeta}}^{\text{II}}-\hat{\mathcal{C}}_{\partial^{j}\partial^{2}\zeta\partial_{j}\dot{\zeta}\partial^{2}\dot{\zeta}}^{\text{II}}+\hat{\mathcal{C}}_{\partial^{4}\zeta\partial^{2}\dot{\zeta}\dot{\zeta}}^{\text{II}} & =0, \label{cond_IIt2s6_4}\\
	-\hat{\mathcal{C}}_{\partial^{2}\zeta\partial^{2}\dot{\zeta}\partial^{2}\dot{\zeta}}^{\text{II}}+\frac{1}{2}\hat{\mathcal{C}}_{\partial^{i}\partial^{j}\zeta\partial^{k}\partial_{i}\dot{\zeta}\partial_{k}\partial_{j}\dot{\zeta}}^{\text{II}}+\hat{\mathcal{C}}_{\partial^{j}\partial^{2}\zeta\partial_{j}\dot{\zeta}\partial^{2}\dot{\zeta}}^{\text{II}}+\hat{\mathcal{C}}_{\partial^{i}\partial^{j}\partial^{2}\zeta\partial_{i}\partial_{j}\dot{\zeta}\dot{\zeta}}^{\text{II}} & =0, \label{cond_IIt2s6_5}
\end{align}
which must hold in order to eliminate the scalar mode. 
After substituting the solutions obtained above, the first two combinations, (\ref{cond_IIt2s6_1}) and (\ref{cond_IIt2s6_2}), vanish identically, while the third condition (\ref{cond_IIt2s6_3}) yields three equations corresponding to different powers of the Hubble constant $H$. Together with the last two conditions, (\ref{cond_IIt2s6_4}) and (\ref{cond_IIt2s6_5}), this gives five equations in total. However, after substituting the previously obtained solutions, one of these five equations becomes inconsistent. We therefore conclude that it is not possible to accommodate a theory with only 2 DOFs in Case II with $\tilde{f}_{3} \neq 0$.

\subsection{$\tilde{f}_{3} = 0$}

We next consider the subcase with $\tilde{f}_{3}=0$.

Under this assumption, the solutions for $A$ and $B$ obtained from (\ref{sol_caseII_A}) and (\ref{sol_caseII_B}) become local, namely,
\begin{align}
	A & =\frac{1+C_{2}\bar{N}}{C_{2}H\bar{N}}\dot{\zeta}-\frac{(d_{8}+2J_{3})(1+C_{2}\bar{N})^{3}}{3C_{2}H\bar{N}^{2}(C_{1}+C_{1}C_{2}\bar{N}+3D_{1}H\bar{N})a^{2}}\partial^{2}\zeta,
\end{align}
and
\begin{align}
	B & =\frac{f_{3}(1+C_{2}\bar{N})^{3}}{2C_{2}H\bar{N}^{2}(C_{1}+C_{1}C_{2}\bar{N}+3D_{1}H\bar{N})a}\dot{\zeta}\nonumber \\
	& \ \ \ \ -\frac{f_{3}(d_{8}+2J_{3})(1+C_{2}\bar{N})^{5}}{3C_{2}H\bar{N}^{3}(C_{1}+C_{1}C_{2}\bar{N}+3D_{1}H\bar{N})^{2}a^{3}}\partial^{2}\zeta\nonumber \\
	& \ \ \ \ +\frac{(1+C_{2}\bar{N})^{2}[C_{2}H(d_{8}+2J_{3})\bar{N}+2(h_{2}+h_{2}^{\prime}+HJ_{3}^{\prime})(1+C_{2}\bar{N})]}{C_{2}H\bar{N}^{2}(C_{1}+C_{1}C_{2}\bar{N}+3D_{1}H\bar{N})a}\zeta.
\end{align}
We again substitute these expressions into the cubic action for the perturbations and classify the resulting terms by the number of temporal derivatives.

\subsubsection{Terms with three temporal derivatives}

The terms with three temporal derivatives take the form
\begin{align}
	\mathcal{L}_{3,t3}^{\text{II}} & \simeq\hat{\mathcal{C}}_{\partial^{j}\partial^{i}\dot{\zeta}\partial_{k}\partial_{i}\dot{\zeta}\partial^{k}\partial_{j}\dot{\zeta}}^{\text{II}}\partial^{j}\partial^{i}\dot{\zeta}\partial_{k}\partial_{i}\dot{\zeta}\partial^{k}\partial_{j}\dot{\zeta}+\hat{\mathcal{C}}_{\partial^{2}\dot{\zeta}\partial^{2}\dot{\zeta}\partial^{2}\dot{\zeta}}^{\text{II}}\partial^{2}\dot{\zeta}\partial^{2}\dot{\zeta}\partial^{2}\dot{\zeta}\nonumber \\
	& \ \ \ \ +\hat{\mathcal{C}}_{\partial^{2}\dot{\zeta}\partial^{i}\partial^{j}\dot{\zeta}\partial_{i}\partial_{j}\dot{\zeta}}^{\text{II}}\partial^{2}\dot{\zeta}\partial^{i}\partial^{j}\dot{\zeta}\partial_{i}\partial_{j}\dot{\zeta}\nonumber \\
	& \ \ \ \ +\hat{\mathcal{C}}_{\partial^{i}\dot{\zeta}\partial_{j}\partial_{i}\dot{\zeta}\partial^{j}\dot{\zeta}}^{\text{II}}\partial^{i}\dot{\zeta}\partial^{j}\dot{\zeta}\partial_{j}\partial_{i}\dot{\zeta}+\hat{\mathcal{C}}_{\partial^{2}\dot{\zeta}\partial^{i}\dot{\zeta}\partial_{i}\dot{\zeta}}^{\text{II}}\partial^{2}\dot{\zeta}\partial^{i}\dot{\zeta}\partial_{i}\dot{\zeta}\nonumber \\
	& \ \ \ \ +\hat{\mathcal{C}}_{\partial^{2}\dot{\zeta}\partial^{2}\dot{\zeta}\dot{\zeta}}^{\text{II}}\partial^{2}\dot{\zeta}\partial^{2}\dot{\zeta}\dot{\zeta}+\hat{\mathcal{C}}_{\partial^{i}\partial^{j}\dot{\zeta}\partial_{i}\partial_{j}\dot{\zeta}\dot{\zeta}}^{\text{II}}\partial^{i}\partial^{j}\dot{\zeta}\partial_{i}\partial_{j}\dot{\zeta}\dot{\zeta}\nonumber \\
	& \ \ \ \ +\hat{\mathcal{C}}_{\partial^{2}\dot{\zeta}\dot{\zeta}\dot{\zeta}}^{\text{II}}\partial^{2}\dot{\zeta}\dot{\zeta}\dot{\zeta}+\hat{\mathcal{C}}_{\partial^{i}\dot{\zeta}\partial_{i}\dot{\zeta}\dot{\zeta}}^{\text{II}}\partial^{i}\dot{\zeta}\partial_{i}\dot{\zeta}\dot{\zeta}.
\end{align}
As before, the terms must be grouped according to the number of spatial derivatives.

For terms with six spatial derivatives, one of the three terms can be reduced by integration by parts, leaving the two conditions
\begin{align}
	-\frac{1}{2}\hat{\mathcal{C}}_{\partial^{j}\partial^{i}\dot{\zeta}\partial_{k}\partial_{i}\dot{\zeta}\partial^{k}\partial_{j}\dot{\zeta}}^{\text{II}}+\hat{\mathcal{C}}_{\partial^{2}\dot{\zeta}\partial^{2}\dot{\zeta}\partial^{2}\dot{\zeta}}^{\text{II}} & =0,\\
	\frac{3}{2}\hat{\mathcal{C}}_{\partial^{j}\partial^{i}\dot{\zeta}\partial_{k}\partial_{i}\dot{\zeta}\partial^{k}\partial_{j}\dot{\zeta}}^{\text{II}}+\hat{\mathcal{C}}_{\partial^{2}\dot{\zeta}\partial^{i}\partial^{j}\dot{\zeta}\partial_{i}\partial_{j}\dot{\zeta}}^{\text{II}} & =0,
\end{align}
which yield
\begin{align}
	c_{3}^{(0;3,0)} & =-\frac{3D_{1}N^{2}}{2(1+C_{2}N)^{2}}=-\frac{3}{2}b_{3}=-\frac{3}{2}\frac{D_{1}}{C_{1}^{2}}b_{2}^{2}. \label{sol_c030_3_caseII}
\end{align}

For terms with four spatial derivatives, two of the four terms can be reduced by integrations by parts, leaving two conditions:
\begin{align}
	-\frac{1}{2}\hat{\mathcal{C}}_{\partial^{i}\dot{\zeta}\partial_{j}\partial_{i}\dot{\zeta}\partial^{j}\dot{\zeta}}^{\text{II}}+\hat{\mathcal{C}}_{\partial^{2}\dot{\zeta}\partial^{i}\dot{\zeta}\partial_{i}\dot{\zeta}}^{\text{II}}+\frac{3}{2}\hat{\mathcal{C}}_{\partial^{i}\partial^{j}\dot{\zeta}\partial_{i}\partial_{j}\dot{\zeta}\dot{\zeta}}^{\text{II}} & =0,\nonumber \\
	\hat{\mathcal{C}}_{\partial^{i}\partial^{j}\dot{\zeta}\partial_{i}\partial_{j}\dot{\zeta}\dot{\zeta}}^{\text{II}}+\hat{\mathcal{C}}_{\partial^{2}\dot{\zeta}\partial^{2}\dot{\zeta}\dot{\zeta}}^{\text{II}} & =0.
\end{align}
The latter vanishes identically. From the first condition, we solve for $c_{4}^{(0;3,0)}$:
\begin{align}
	c_{4}^{(0;3,0)} & =\frac{7}{20}\frac{C_{2}N}{1+C_{2}N}f_{3}+\frac{1}{2}f_{3}^{\prime}, \label{sol_c030_4_caseII}
\end{align}
or else $c_{4}^{(0;3,0)}=0$ with $f_{3}=C_{6}/(1+C_{2}N)^{7/10}$. 

For terms with two spatial derivatives, one of the two terms can be reduced by integration by parts, leaving the single condition
\begin{align}
	-\frac{1}{2}\hat{\mathcal{C}}_{\partial^{i}\dot{\zeta}\partial_{i}\dot{\zeta}\dot{\zeta}}^{\text{II}}+\hat{\mathcal{C}}_{\partial^{2}\dot{\zeta}\dot{\zeta}\dot{\zeta}}^{\text{II}}= & 0.
\end{align}
This determines $f_{3}$ as
\begin{align}
	f_{3} & =D_{6}.
\end{align}
After imposing this condition, the solution (\ref{sol_c030_4_caseII}) reduces to
\begin{align}
	c_{4}^{(0;3,0)} & =\frac{7}{20}\frac{C_{2}N}{1+C_{2}N}D_{6}.
\end{align}

\subsubsection{Terms with two temporal derivatives}

We next consider the terms in the cubic Lagrangian involving two temporal derivatives:
\begin{align}
	\mathcal{L}_{3,t2}^{\text{II}} & \simeq\hat{\mathcal{C}}_{\partial^{i}\partial^{j}\partial^{2}\zeta\partial_{i}\partial_{j}\dot{\zeta}\partial^{2}\dot{\zeta}}^{\text{II}}\partial^{i}\partial^{j}\partial^{2}\zeta\partial_{i}\partial_{j}\dot{\zeta}\partial^{2}\dot{\zeta}+\hat{\mathcal{C}}_{\partial^{i}\partial^{j}\partial^{2}\zeta\partial^{k}\partial_{i}\dot{\zeta}\partial_{k}\partial_{j}\dot{\zeta}}^{\text{II}}\partial^{i}\partial^{j}\partial^{2}\zeta\partial^{k}\partial_{i}\dot{\zeta}\partial_{k}\partial_{j}\dot{\zeta}\nonumber \\
	& \ \ \ \ +\hat{\mathcal{C}}_{\partial^{4}\zeta\partial^{j}\partial^{i}\dot{\zeta}\partial_{i}\partial_{j}\dot{\zeta}}^{\text{II}}\partial^{4}\zeta\partial^{j}\partial^{i}\dot{\zeta}\partial_{i}\partial_{j}\dot{\zeta}+\hat{\mathcal{C}}_{\partial^{4}\zeta\partial^{2}\dot{\zeta}\partial^{2}\dot{\zeta}}^{\text{II}}\partial^{4}\zeta\partial^{2}\dot{\zeta}\partial^{2}\dot{\zeta}\nonumber \\
	& \ \ \ \ +\hat{\mathcal{C}}_{\partial^{i}\partial^{j}\zeta\partial^{k}\partial_{i}\dot{\zeta}\partial_{k}\partial_{j}\dot{\zeta}}^{\text{II}}\partial^{i}\partial^{j}\zeta\partial^{k}\partial_{i}\dot{\zeta}\partial_{k}\partial_{j}\dot{\zeta}+\hat{\mathcal{C}}_{\partial^{i}\partial^{j}\zeta\partial_{i}\partial_{j}\dot{\zeta}\partial^{2}\dot{\zeta}}^{\text{II}}\partial^{i}\partial^{j}\zeta\partial_{i}\partial_{j}\dot{\zeta}\partial^{2}\dot{\zeta}\nonumber \\
	& \ \ \ \ +\hat{\mathcal{C}}_{\partial^{2}\zeta\partial^{j}\partial^{i}\dot{\zeta}\partial_{i}\partial_{j}\dot{\zeta}}^{\text{II}}\partial^{2}\zeta\partial^{j}\partial^{i}\dot{\zeta}\partial_{i}\partial_{j}\dot{\zeta}+\hat{\mathcal{C}}_{\partial^{2}\zeta\partial^{2}\dot{\zeta}\partial^{2}\dot{\zeta}}^{\text{II}}\partial^{2}\zeta\partial^{2}\dot{\zeta}\partial^{2}\dot{\zeta}\nonumber \\
	& \ \ \ \ +\hat{\mathcal{C}}_{\partial^{i}\partial^{j}\partial^{2}\zeta\partial_{i}\dot{\zeta}\partial_{j}\dot{\zeta}}^{\text{II}}\partial^{i}\partial^{j}\partial^{2}\zeta\partial_{i}\dot{\zeta}\partial_{j}\dot{\zeta}+\hat{\mathcal{C}}_{\partial^{j}\partial^{2}\zeta\partial_{k}\partial_{j}\dot{\zeta}\partial^{k}\dot{\zeta}}^{\text{II}}\partial^{j}\partial^{2}\zeta\partial_{k}\partial_{j}\dot{\zeta}\partial^{k}\dot{\zeta}\nonumber \\
	& \ \ \ \ +\hat{\mathcal{C}}_{\partial^{4}\zeta\partial^{2}\dot{\zeta}\dot{\zeta}}^{\text{II}}\partial^{4}\zeta\partial^{2}\dot{\zeta}\dot{\zeta}+\hat{\mathcal{C}}_{\partial^{i}\partial^{j}\partial^{2}\zeta\partial_{i}\partial_{j}\dot{\zeta}\dot{\zeta}}^{\text{II}}\partial^{i}\partial^{j}\partial^{2}\zeta\partial_{i}\partial_{j}\dot{\zeta}\dot{\zeta}\nonumber \\
	& \ \ \ \ +\hat{\mathcal{C}}_{\partial^{j}\partial^{2}\zeta\partial_{j}\dot{\zeta}\partial^{2}\dot{\zeta}}^{\text{II}}\partial^{j}\partial^{2}\zeta\partial_{j}\dot{\zeta}\partial^{2}\dot{\zeta}+\hat{\mathcal{C}}_{\partial^{4}\zeta\partial_{i}\dot{\zeta}\partial^{i}\dot{\zeta}}^{\text{II}}\partial^{4}\zeta\partial_{i}\dot{\zeta}\partial^{i}\dot{\zeta}\nonumber \\
	& \ \ \ \ +\hat{\mathcal{C}}_{\zeta\partial^{j}\partial^{i}\dot{\zeta}\partial_{i}\partial_{j}\dot{\zeta}}^{\text{II}}\zeta\partial^{j}\partial^{i}\dot{\zeta}\partial_{i}\partial_{j}\dot{\zeta}+\hat{\mathcal{C}}_{\zeta\partial^{2}\dot{\zeta}\partial^{2}\dot{\zeta}}^{\text{II}}\zeta\partial^{2}\dot{\zeta}\partial^{2}\dot{\zeta}+\hat{\mathcal{C}}_{\partial^{j}\partial^{i}\zeta\partial_{i}\dot{\zeta}\partial_{j}\dot{\zeta}}^{\text{II}}\partial^{j}\partial^{i}\zeta\partial_{i}\dot{\zeta}\partial_{j}\dot{\zeta}\nonumber \\
	& \ \ \ \ +\hat{\mathcal{C}}_{\partial^{i}\zeta\partial^{j}\dot{\zeta}\partial_{i}\partial_{j}\dot{\zeta}}^{\text{II}}\partial^{i}\zeta\partial^{j}\dot{\zeta}\partial_{i}\partial_{j}\dot{\zeta}+\hat{\mathcal{C}}_{\partial^{2}\zeta\partial^{i}\dot{\zeta}\partial_{i}\dot{\zeta}}^{\text{II}}\partial^{2}\zeta\partial^{i}\dot{\zeta}\partial_{i}\dot{\zeta}+\hat{\mathcal{C}}_{\partial^{2}\zeta\partial^{2}\dot{\zeta}\dot{\zeta}}^{\text{II}}\partial^{2}\zeta\partial^{2}\dot{\zeta}\dot{\zeta}\nonumber \\
	& \ \ \ \ +\hat{\mathcal{C}}_{\partial^{i}\partial^{j}\zeta\partial_{i}\partial_{j}\dot{\zeta}\dot{\zeta}}^{\text{II}}\partial^{i}\partial^{j}\zeta\partial_{i}\partial_{j}\dot{\zeta}\dot{\zeta}+\hat{\mathcal{C}}_{\partial^{i}\partial^{2}\zeta\partial_{i}\dot{\zeta}\dot{\zeta}}^{\text{II}}\partial^{i}\partial^{2}\zeta\partial_{i}\dot{\zeta}\dot{\zeta}+\hat{\mathcal{C}}_{\partial^{4}\zeta\dot{\zeta}\dot{\zeta}}^{\text{II}}\partial^{4}\zeta\dot{\zeta}^{2}\nonumber \\
	& \ \ \ \ +\hat{\mathcal{C}}_{\zeta\partial^{i}\dot{\zeta}\partial_{i}\dot{\zeta}}^{\text{II}}\zeta\partial^{i}\dot{\zeta}\partial_{i}\dot{\zeta}+\hat{\mathcal{C}}_{\zeta\partial^{2}\dot{\zeta}\dot{\zeta}}^{\text{II}}\zeta\partial^{2}\dot{\zeta}\dot{\zeta}+\hat{\mathcal{C}}_{\partial^{2}\zeta\dot{\zeta}\dot{\zeta}}^{\text{II}}\partial^{2}\zeta\dot{\zeta}^{2}.
\end{align}

For terms with eight spatial derivatives, one of the four terms can be reduced by integration by parts, leaving three conditions:
\begin{align}
	\hat{\mathcal{C}}_{\partial^{i}\partial^{j}\partial^{2}\zeta\partial_{i}\partial_{j}\dot{\zeta}\partial^{2}\dot{\zeta}}+\hat{\mathcal{C}}_{\partial^{i}\partial^{j}\partial^{2}\zeta\partial^{k}\partial_{i}\dot{\zeta}\partial_{k}\partial_{j}\dot{\zeta}}^{\text{II}} & =0,\\
	\frac{1}{2}\hat{\mathcal{C}}_{\partial^{i}\partial^{j}\partial^{2}\zeta\partial_{i}\partial_{j}\dot{\zeta}\partial^{2}\dot{\zeta}}^{\text{II}}+\hat{\mathcal{C}}_{\partial^{4}\zeta\partial^{2}\dot{\zeta}\partial^{2}\dot{\zeta}}^{\text{II}} & =0,\\
	-\frac{1}{2}\hat{\mathcal{C}}_{\partial^{i}\partial^{j}\partial^{2}\zeta\partial_{i}\partial_{j}\dot{\zeta}\partial^{2}\dot{\zeta}}^{\text{II}}+\hat{\mathcal{C}}_{\partial^{4}\zeta\partial^{j}\partial^{i}\dot{\zeta}\partial_{i}\partial_{j}\dot{\zeta}}^{\text{II}} & =0.
\end{align}
The second condition is trivial, and the other two reproduce the same solution as (\ref{sol_c030_3_caseII}), namely,
\begin{align}
	c_{3}^{(0;3,0)} & =-\frac{3D_{1}N^{2}}{2(1+C_{2}N)^{2}}=-\frac{3}{2}b_{3}=-\frac{3}{2}\frac{D_{1}}{C_{1}^{2}}b_{2}^{2}.
\end{align}

For terms with six spatial derivatives, five of the ten terms can be reduced by integrations by parts, leaving five conditions:
\begin{align}
	\hat{\mathcal{C}}_{\partial^{i}\partial^{j}\zeta\partial^{k}\partial_{i}\dot{\zeta}\partial_{k}\partial_{j}\dot{\zeta}}^{\text{II}}+\hat{\mathcal{C}}_{\partial^{i}\partial^{j}\zeta\partial_{i}\partial_{j}\dot{\zeta}\partial^{2}\dot{\zeta}}^{\text{II}} & =0,\\
	\hat{\mathcal{C}}_{\partial^{2}\zeta\partial^{j}\partial^{i}\dot{\zeta}\partial_{i}\partial_{j}\dot{\zeta}}^{\text{II}}+\hat{\mathcal{C}}_{\partial^{2}\zeta\partial^{2}\dot{\zeta}\partial^{2}\dot{\zeta}}^{\text{II}} & =0,\\
	\hat{\mathcal{C}}_{\partial^{4}\zeta\partial^{2}\dot{\zeta}\dot{\zeta}}^{\text{II}}+\hat{\mathcal{C}}_{\partial^{i}\partial^{j}\partial^{2}\zeta\partial_{i}\partial_{j}\dot{\zeta}\dot{\zeta}}^{\text{II}} & =0,\\
	\frac{1}{2}\hat{\mathcal{C}}_{\partial^{i}\partial^{j}\zeta\partial^{k}\partial_{i}\dot{\zeta}\partial_{k}\partial_{j}\dot{\zeta}}^{\text{II}}+\hat{\mathcal{C}}_{\partial^{2}\zeta\partial^{j}\partial^{i}\dot{\zeta}\partial_{i}\partial_{j}\dot{\zeta}}^{\text{II}}-\hat{\mathcal{C}}_{\partial^{i}\partial^{j}\partial^{2}\zeta\partial_{i}\dot{\zeta}\partial_{j}\dot{\zeta}}^{\text{II}}-\hat{\mathcal{C}}_{\partial^{4}\zeta\partial^{2}\dot{\zeta}\dot{\zeta}}^{\text{II}}+\hat{\mathcal{C}}_{\partial^{j}\partial^{2}\zeta\partial_{j}\dot{\zeta}\partial^{2}\dot{\zeta}}^{\text{II}} & =0,\\
	\frac{1}{4}\hat{\mathcal{C}}_{\partial^{i}\partial^{j}\zeta\partial^{k}\partial_{i}\dot{\zeta}\partial_{k}\partial_{j}\dot{\zeta}}^{\text{II}}+\frac{1}{2}\hat{\mathcal{C}}_{\partial^{2}\zeta\partial^{j}\partial^{i}\dot{\zeta}\partial_{i}\partial_{j}\dot{\zeta}}^{\text{II}}+\frac{1}{2}\hat{\mathcal{C}}_{\partial^{i}\partial^{j}\partial^{2}\zeta\partial_{i}\dot{\zeta}\partial_{j}\dot{\zeta}}^{\text{II}}\ \ \ \ \nonumber \\
	-\frac{1}{2}\hat{\mathcal{C}}_{\partial^{j}\partial^{2}\zeta\partial_{k}\partial_{j}\dot{\zeta}\partial^{k}\dot{\zeta}}^{\text{II}}-\frac{1}{2}\hat{\mathcal{C}}_{\partial^{4}\zeta\partial^{2}\dot{\zeta}\dot{\zeta}}^{\text{II}}+\hat{\mathcal{C}}_{\partial^{4}\zeta\partial_{i}\dot{\zeta}\partial^{i}\dot{\zeta}}^{\text{II}} & =0.
\end{align}
The first three conditions vanish after substituting the solutions obtained above, while the remaining two imply the same result,
\begin{align}
	c_{1}^{(1;1,0)} & =-2J_{3}.
\end{align}

For terms with four spatial derivatives, five of the nine terms can be reduced by integrations by parts, leaving four conditions:
\begin{align}
	-\hat{\mathcal{C}}_{\zeta\partial^{j}\partial^{i}\dot{\zeta}\partial_{i}\partial_{j}\dot{\zeta}}^{\text{II}}+\frac{1}{2}\hat{\mathcal{C}}_{\partial^{i}\zeta\partial^{j}\dot{\zeta}\partial_{i}\partial_{j}\dot{\zeta}}^{\text{II}}-\hat{\mathcal{C}}_{\partial^{2}\zeta\partial^{i}\dot{\zeta}\partial_{i}\dot{\zeta}}^{\text{II}}+\hat{\mathcal{C}}_{\partial^{2}\zeta\partial^{2}\dot{\zeta}\dot{\zeta}}^{\text{II}}= & 0,\\
	\hat{\mathcal{C}}_{\zeta\partial^{j}\partial^{i}\dot{\zeta}\partial_{i}\partial_{j}\dot{\zeta}}^{\text{II}}-\hat{\mathcal{C}}_{\partial^{j}\partial^{i}\zeta\partial_{i}\dot{\zeta}\partial_{j}\dot{\zeta}}^{\text{II}}+\hat{\mathcal{C}}_{\partial^{i}\partial^{j}\zeta\partial_{i}\partial_{j}\dot{\zeta}\dot{\zeta}}^{\text{II}}= & 0,\\
	\frac{1}{2}\hat{\mathcal{C}}_{\partial^{j}\partial^{i}\zeta\partial_{i}\dot{\zeta}\partial_{j}\dot{\zeta}}^{\text{II}}-\frac{1}{4}\hat{\mathcal{C}}_{\partial^{i}\zeta\partial^{j}\dot{\zeta}\partial_{i}\partial_{j}\dot{\zeta}}^{\text{II}}+\frac{1}{2}\hat{\mathcal{C}}_{\partial^{2}\zeta\partial^{i}\dot{\zeta}\partial_{i}\dot{\zeta}}^{\text{II}}-\frac{1}{2}\hat{\mathcal{C}}_{\partial^{i}\partial^{2}\zeta\partial_{i}\dot{\zeta}\dot{\zeta}}^{\text{II}}+\hat{\mathcal{C}}_{\partial^{4}\zeta\dot{\zeta}\dot{\zeta}}^{\text{II}}= & 0,\\
	\hat{\mathcal{C}}_{\zeta\partial^{j}\partial^{i}\dot{\zeta}\partial_{i}\partial_{j}\dot{\zeta}}^{\text{II}}+\hat{\mathcal{C}}_{\zeta\partial^{2}\dot{\zeta}\partial^{2}\dot{\zeta}}^{\text{II}}= & 0.
\end{align}
Although the last condition vanishes identically, the others are satisfied only if $f_{3}=0$. This, however, is incompatible with the basic assumption of Case II; see (\ref{case2omg2}) and (\ref{eq:solution for omega3}). 
Hence, we conclude that it is not possible to obtain coefficient solutions with 2 DOFs in Case II with $\tilde{f}_{3}=0$.

\bibliographystyle{JHEP}
\bibliography{ref}

\end{document}